**Advancements and challenges in plasmon-exciton quantum emitters based on colloidal quantum dots**

Adam Olejniczak,[1] Yury Rakovich,[1,2,3,4], Victor Krivenkov[1,4]*

[1]Centro de Física de Materiales (MPC, CSIC-UPV/EHU), Donostia - San Sebastián, 20018, Spain

[2]Donostia International Physics Center (DIPC), Donostia - San Sebastián, 20018, Spain

[3]Ikerbasque, Basque Foundation for Science, Bilbao, 48013, Spain

[4]Polymers and Materials: Physics, Chemistry and Technology, Chemistry Faculty, University of the Basque Country (UPV/EHU), Donostia - San Sebastián, 20018, Spain

*victor.krivenkov@ehu.eus

The Nobel Prizes in Physics (2022) and Chemistry (2023) heralded the recognition of quantum information science and the synthesis of quantum dots, respectively. This acknowledgment has propelled colloidal quantum dots and perovskite nanocrystals to the forefront of quantum technologies. Their distinct emission properties, facilitating the efficient generation of both single photons and photon pairs, render them particularly captivating. Moreover, their adaptability to diverse structures, ranging from traditional electronics to nanopatterned frameworks, underscores their pivotal role in shaping quantum technologies. Despite notable strides in synthesis, certain properties require refinement for enhanced applicability in quantum information, encompassing emission brightness, stability, single-photon indistinguishability, and entanglement fidelity of photon pairs. Here we offer an overview of recent achievements in plasmon-exciton quantum emitters based on luminescent semiconductor nanocrystals. Emphasizing the utilization of the light-matter coupling phenomenon, we explore how this interaction enables the manipulation of quantum properties without altering the chemical structure of the emitters. This approach addresses critical aspects for quantum information applications, offering precise control over emission rate, intensity, and energy. The development of these hybrid systems represents a significant stride forward, demonstrating their potential to overcome existing challenges and advance the integration of quantum emitters into cutting-edge quantum technology applications.

## Introduction

In the past three decades, the field of quantum information has transitioned from predominantly basic research to applied research, with a strong focus on developing new platforms for quantum communication and quantum computation. Over the last 15 years, practical quantum computation systems have been successfully demonstrated, showcasing their capability to solve technologically relevant algorithmic problems in the near future [1]. At the heart of quantum information is the quantum bit (qubit), fundamentally distinct from its classical counterpart. Classical bits are limited to values of 0 or 1, whereas qubits can exist in a superposition of 0 and 1 simultaneously. The unique quantum nature of qubits enables provably secure communication, long-term secure storage, cloud computing, and other cryptography-related tasks. In the future, this may lead to the establishment of a secure 'quantum internet' [2,3]. The secure transfer of photons over long distances with minimal losses has advanced considerably, reaching distances of hundreds of kilometers. Networks built upon these connections have demonstrated spans of over 4600 kilometers [4,5]. Ensuring the safety of quantum communication involves using only a single photon for each encoded pulse, guarding against potential splitting attacks [6]. Therefore, the ideal light source for quantum information should exhibit single-photon statistics, commonly referred to as antibunching. The measurement of photon statistics can be accomplished using the Hanbury Brown and Twiss (HBT) geometry of the experiment (Figure 1a, b). In this setup, the signal from the photon source is split and directed towards two photodetectors. The second-order cross-correlation function $g^{(2)}$ can be employed to estimate the type of photon statistics exhibited by the light source:

$$g^{(2)}(\tau) = \frac{\langle I_1(t) I_2(t+\tau)\rangle}{\langle I_1(t)\rangle\langle I_2(t+\tau)\rangle}, (1)$$

where $I_i(t)$ is a photon detection event on one of two detectors, $\tau$ is a time delay between two-photon detection events and $\langle\cdots\rangle$ is a time averaging. The pure antibunching is evidenced by the zero value of the $g^{(2)}$ at $\tau = 0$, which represents the pure single-photon behavior of the emitter. In the case of continuous wave (CW) excitation, antibunching manifests itself as a dip at τ = 0 (red line in Figure 1c, d). Under pulsed excitation the $g^{(2)}$ curve consists of a series of discrete peaks separated at the same distance as laser pulses. The pure antibunching relates to the zero central peak but the single-photon emission in general is related to the level of $g^{(2)}(\tau=0)$ less than 0.5 (blue line in Figure 1c) [7]. The multiphoton emission is related to the level of $g^{(2)}(\tau=0)$ more than 0.5 (Figure 1d) [7].

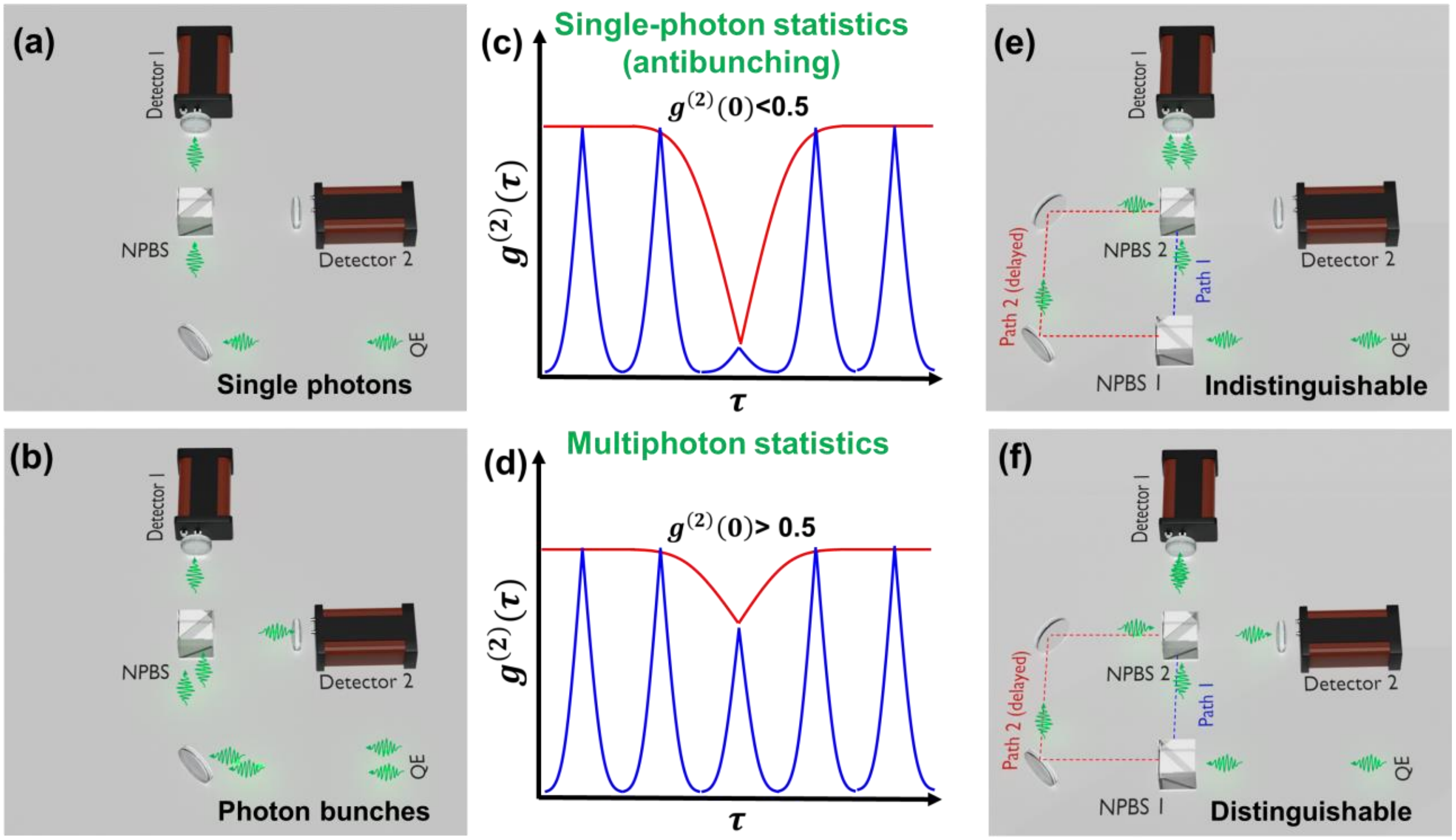

**Figure 1.** Photon correlation techniques commonly used to determine emission statistics type. (a, b) HBT experiment with (a) a stream of single photons and (b) a stream of photon bunches. (c, d) characteristic second order cross-correlation function $g^{(2)}$ (red curve – CW excitation, blue curve – pulsed excitation) measured in HBT setup for (c) single-photon QE, and (d) multiphoton emitter. (e, f) HOM experiment with Mach–Zehnder interferometer in the case of a stream of (e) indistinguishable single photons, and (f) distinguishable single photons.

The optimal single-photon source, commonly referred to as a quantum emitter (QE), is designed to emit precisely one photon on-demand (at a time defined by the user) into a specified spatiotemporal mode [8]. Additionally, photons emitted by the ideal QE are indistinguishable, exhibiting full interference and coherence, ensuring their quantum properties are identical. The Hong–Ou–Mandel (HOM) effect is commonly employed to measure indistinguishability. In the HOM measurement geometry, two photons arrive at the non-polarizing beamsplitter (NPBS) from different arms (Figure 1e,f). In simple terms, zero coincidences between two detectors at zero time delay between two photons at the beam splitter indicate their indistinguishability (Figure 1e). The primary parameter that reflects the level of indistinguishability is the HOM visibility ($V_{HOM}$), which can be calculated by measuring the central areas in the correlation functions of spatially indistinguishable ($A$) and distinguishable ($A'$) photons: $V_{HOM} = (A' - A)/A'$ [9].

Several promising quantum information protocols involve not only indistinguishable single photons but also quantum entangled photon pairs. In the context of quantum informatics, quantum entanglement implies that measuring the state of one entangled photon simultaneously provides information about the state of the second photon. The generation of single entangled photon pairs is even more challenging than single indistinguishable photons due to limited available sources. [10,11].

Indistinguishable single photons and entangled photon pairs have traditionally been generated through parametric down-conversion [12]. This nonlinear process yields a probabilistic amount of photons, leading to the generation of zero and multiple photons/photon pairs. Consequently, it cannot reliably produce single entangled photon/photon pairs on demand, rendering them unsuitable as QEs. Recent efforts in QE design have shifted towards cold atoms and ions, superconducting circuits, solid-state sources like emitting vacancy centers in diamonds, and semiconductor based QEs [13,14]. Among these, semiconductor-based QEs offer greater flexibility for integration into optical logic schemes and combination with traditional silicon-based electronics. Strongly and moderately confined luminescent semiconductor nanocrystals can serve as QEs [15]. Semiconductor QEs can be formed in semiconducting layered materials as self-assembled quantum dots (QDs), strain-induced QDs, or epitaxial QDs [16], or through imperfections in 2D materials, predominantly transition metal dichalcogenides (TMDC) or hexagonal boron-nitride [17]. Additionally, semiconductor QEs can be chemically synthesized in the form of colloidal nanoparticles, with colloidal semiconductor QDs (CQDs) and perovskite nanocrystals (PNCs) being noteworthy examples. Semiconductor QEs are capable of emitting both indistinguishable single photons and entangled photon pairs [14]. Compared to other solid-state QEs, they exhibit superior single-photon purity and a narrower emission spectrum [8,14].

**Overview of the current stage of colloidal semiconductor quantum emitters**

Perovskite nanocrystals (PNCs) and colloidal quantum dots (CQDs) share several common features, including high photoluminescence quantum yield (PL QY), tunable emission, and a possibility to excite them with any wavelength above the bandgap (Figure 2b,e). Since the discovery of colloidal semiconductor nanocrystals in 1980s, wet-chemistry synthesis protocols have undergone significant improvements, enabling the production of highly crystalline nanomaterials with precise control over size and minimal defect density [18] (Figure 2 a,b). Although core-only CQDs have larger volume-normalized absorption cross-sections compared to PNCs, the total photon absorption cross-section of PNCs can be an order of magnitude greater than that of CQDs, exceeding $10^{-13}$ $cm^2$ due to the PNC volume being more than 7000 $nm^3$ [19,20]. Thick-shell CdSe/CdS CQDs, frequently utilized as QEs (refer to Table 1), exhibit absorption cross-section values similar to those of PNCs [21]. For comparison, epitaxial QDs have absorption cross-sections orders of magnitude smaller [22,23]. The substantial absorption cross-sections of both PNCs and CQDs facilitate easy excitation of these colloidal QEs by light, enabling on-demand regimes with lower excitation power than commonly used for epitaxial QDs.

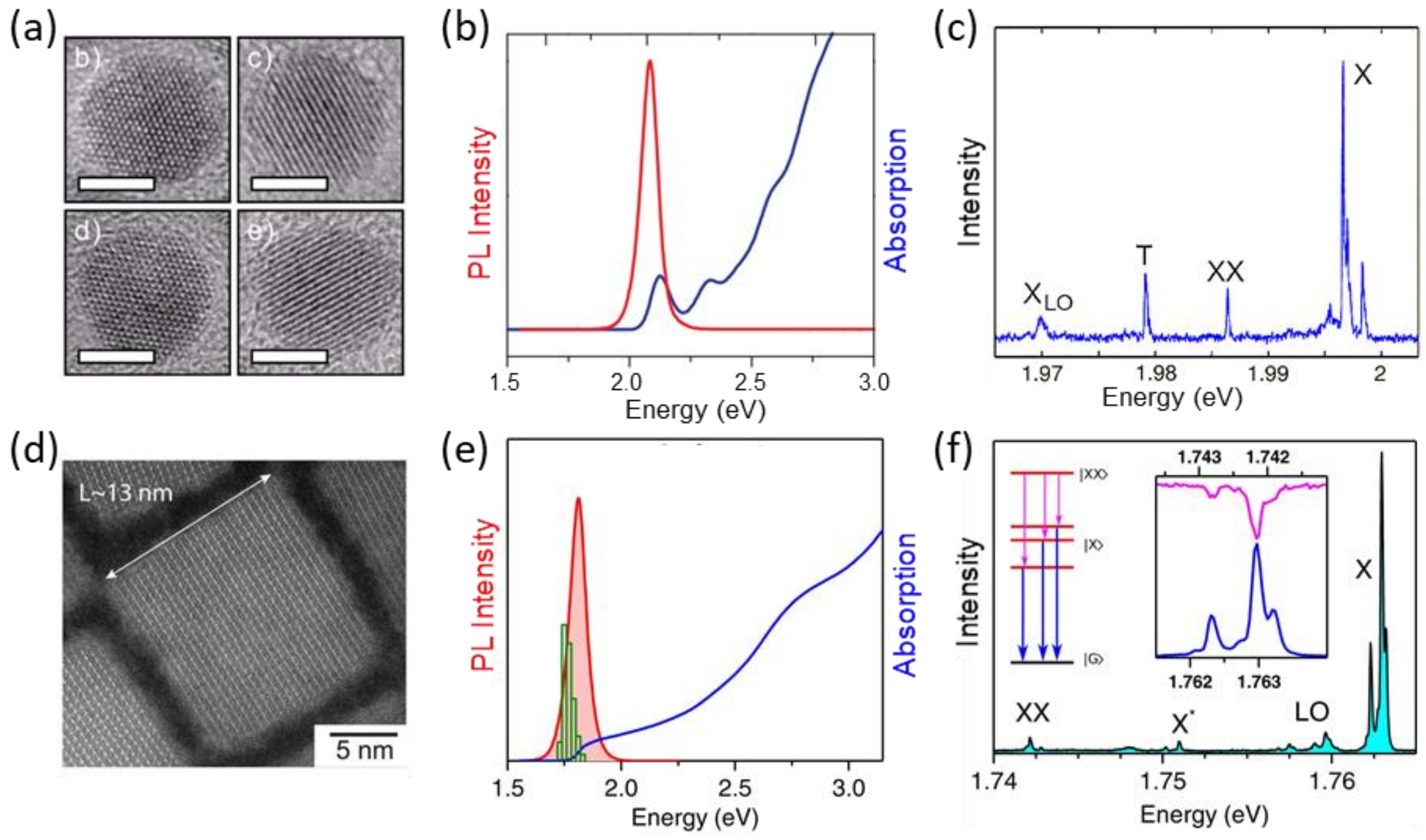

**Figure 2.** Representative examples of structure, room temperature and low temperature spectroscopic properties of (a, b, c) CQDs and (d, e, f) PNCs. (a and d) High-resolution TEM images of CdSe/CdS core/shell CQDs and $CsPbBr_3$ PNC, respectively. Scale bars in (a) are 5 nm. (b and e) Representative room temperature PL (red) and absorption (blue) spectra from ensemble of CdSe/CdS core–shell CQDs and ensemble of $CsPbI_3$ NCs, respectively. The green histogram in (e) displays the distribution of emission energies of 58 single NCs at 4 K. (c and f) Emission spectra of single CdSe/ZnS nanocrystal at 2 K and single $CsPbI_3$ NC at 4 K, respectively. Inset in (f) shows

correspondence between X and XX multiplets. (a, b) Adapted from [24]. (c) Adapted with permission from [25]. Copyright 2024 American Chemical Society. (d) From [26]. Reprinted with permission from AAAS. (e, f) Adapted from [27].

Another noteworthy feature shared by both PNCs and CQDs is the ability to excite more than one electron-hole pair, leading to the formation of multiexciton states in these QEs. The initial multiexciton state is the biexciton state, representing two excited electrons and two holes in one nanocrystal. Biexciton formation may result from either two-photon excitation of the nanocrystal or exciton multiplication during the relaxation of a highly excited carriers with energy higher than two bandgap energies [28]. To achieve complete relaxation of the nanocrystal with an excited biexciton, it must undergo the cascade recombination process from the biexciton state (XX) to the single-exciton state (X) and then to the ground state (GS) (Figure 3a). The photon pair emitted during the XX→X→GS cascade recombination may be quantum entangled by polarization, making nanocrystals promising emitters of entangled photon pairs [29]. Thick-shell CdSe/CdS CQDs exhibit a significant biexciton blueshift of 24 ± 5 nm, comparable to the exciton spectral width (40 ± 5 nm) [30], and the similar values could be achieved for PNCs [31]. This spectral separation allows the spectral isolation of photons produced by the biexciton and exciton (Figure 3a), thereby preserving the single-photon purity of the emission if required [30]. Furthermore, the biexciton emission rate is at least four times higher than that for the exciton, enabling temporal filtering of the signal for additional enhancement of single-photon purity [32].

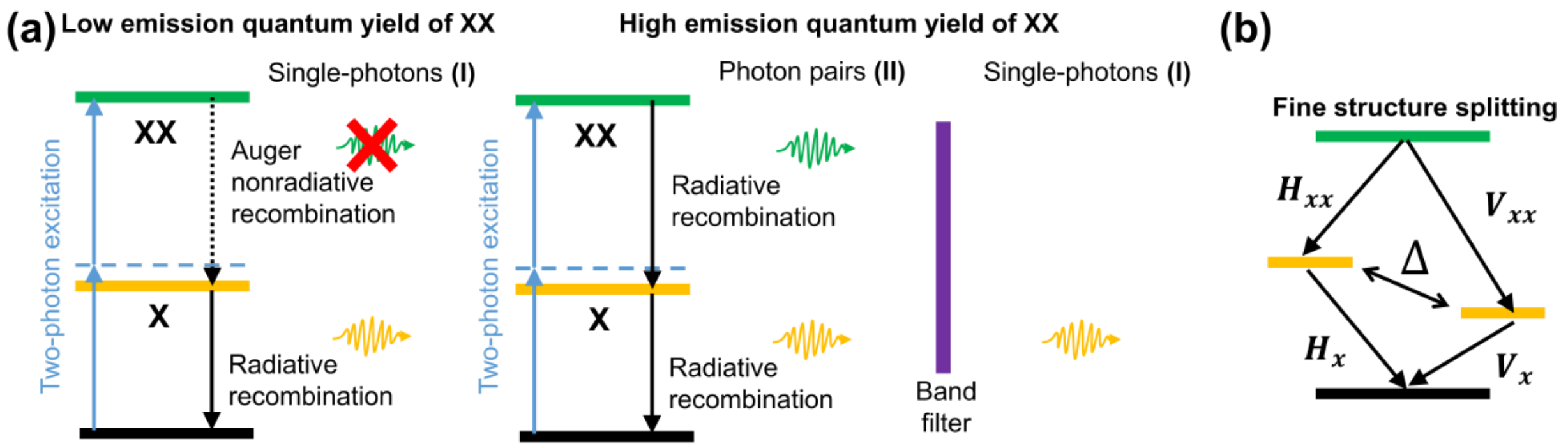

**Figure 3.** (a) Two ways to achieve single-photon emission from semiconductor QE: by keeping low XX PL QY, and by the filtering of the photon produced by XX→X transition. (b) Schematic illustration of the connection between polarization (H or V) and energies of photons emitted by XX→X→GS cascade recombination due FSS (Δ) of the intermediate X level.

Nevertheless, the predominant type of semiconductor QEs in current use is epitaxial QDs. They exhibit the advantage of a narrow PL line at helium temperatures [33], enabling the achievement of indistinguishability in emitted single photons. However, the PL QY of epitaxial QDs is significantly lower than that of colloidal nanocrystals [34]. At temperatures ranging from 50-280 K, the single-photon purity of epitaxial QDs is also comparatively poor [35–37]. The internal reflection within epitaxial QDs limits photon-extraction efficiency to less than 2% for an InAs/GaAs QD [8]. Consequently, recent efforts in QD-based QE fabrication have focused on creating cavities and photonic antennas to enhance emission brightness and increase photon collection efficiency [16,29]. However, achieving cavity control over the emission properties of epitaxial QDs necessitates growing cavity structures simultaneously with QD formation, precluding the substitution of other cavity types. In contrast, leveraging light-matter coupling in external tunable microcavities allows precise control over the emission properties of QEs [38]. The utilization of plasmonic nanocavities, such as bow-tie and nanopatch plasmonic nanoantennas, enables orders of magnitude higher values of the Purcell factor and light-matter coupling strength compared to open plasmonic cavities (single nanoparticles) [39–41]. However, applying them to epitaxial QDs is challenging as plasmonic structures are commonly grown together with the QD [16]. CQDs and PNCs do not suffer from these limitations. They boast near 100% PL QY and low $g^{(2)}(0)$ values even at room temperatures (refer to Table 1). Furthermore, colloidal nanocrystals of the same type and synthesis batch can be coupled to different cavity types [42–44], offering high flexibility in tuning their emission properties [45]. Ordered arrays of colloidal semiconductor QEs can be produced by capillary-assisted deposition without requiring high-precision deposition techniques [45,46]. For these reasons, recent efforts in plasmon-exciton coupled QEs have primarily focused on the use of colloidal semiconductor QEs and replaceable TMDC materials [47]. Therefore, in this review, our main focus is on plasmon-exciton QEs based on colloidal luminescent nanocrystals rather than epitaxial ones. Table 1 summarizes the current stage in the development of colloidal semiconductor QEs of different materials, including essential QE parameters such as emission wavelength, peak width (related to the damping rate), PL QY, single-photon purity ($g^{(2)}(0)$ value), and exciton and biexciton emission lifetimes. For more detailed information, including 2D materials, refer to Tables S1A, S1B and S1C in the Supplementary material file. Colloidal QEs have proven themselves as a potential source of indistinguishable photons [9,48]. In 2023, Proppe et al. demonstrated the potential applicability of InP/ZnSe/ZnS CQDs for the generation of indistinguishable photons [48]. In the same year, Kaplan et al. demonstrated $V_{HOM}$ visibilities of up to 0.56 ± 0.12 for $CsPbBr_3$ PNCs, bringing them closer to achieving indistinguishable photons [9].

**Table. 1.** Different types of single colloidal semiconductor QEs and their emission properties. If the temperature is not specified, the data in the table refers to room temperature.

| Material | PL max | PL FWHM | $g_0^{(2}$ | PL QY | X lifetime | XX lifetime | Ref. |
|---|---|---|---|---|---|---|---|
| CQDs | | | | | | | |
| Core only | | | | | | | |
| CdSe | 605 (RT) | - | 0.33 | 40-50% | 1.7 ns | - | [49] |
| Core-shell | | | | | | | |
| CdSe/ZnS | ~ 570 nm | - | 0.05<br>0.004 (filtered) | 40% | 20 ns | - | [50,51] |
| CdSe/ZnS/CdS/ZnS | ~ 560 nm | 39 nm (ensemble) | 0.12 | ~50% | 15-18 ns | 0.5 ns | [43,44] |
| $CdSe/Zn_{1-x}CdxS$ | ~ 635 nm | ~45 nm (ensemble) | <0.25 | - | 12 ns | - | [34] |
| CdSe/CdS | 620-655 nm | ~65.1 meV | 0.04-0.15<br>0.03 (after spectral filtering) | 50-90% | 60-125 ns, up to 638 ns | 0.7-6.44 ns | [30,52–56] |
| $CdSe/CdS/SiO_2$ | 635-645 nm | ~40 nm (ensemble) | 0.06-0.33 (RT) | 34-47% | ~ 30 ns | - | [46,57] |
| InP/ZnSe/ZnS | 625 nm | 5 μeV (4 K) | 0.13 (RT), 0.07 (4 K) | - | 3 ns, and 16.7 ns (4 K) | - | [48] |
| PbS/CdS | 1280 nm / 1500 nm | 90 meV | 0.4 (RT) | 10-30% | 1000 ns / 3500 ns | - | [58] |
| InP/ZnSe | 629 nm | 40-80 meV | 0.03 (RT) | 70-100% | 21 ns | 0.070 ns | [59] |
| CdSeTe/ZnS | 705 nm | ~100 meV | 0.01-0.06 (RT) | - | 138 ns | 2 ns | [32] |
| CdSe/CdS (dot-in-rod) | 579-605 nm | 100-140 meV | 0.075 (RT)<br>0.11 (RT, bright state), 0.47 (RT, grey state) | 100% (bright state)<br>36% (grey state) | 46 ns<br>65 ns (bright state), 11.6 ns (grey state) | - | [60,61] |
| $CuInS_2/ZnS$ (Tetrahedrona) | ~1.93 eV | 128 meV | ~0.2 (RT) | 50% | 80-328 ns | - | [62] |
| CdSe/CdS (dot-in-tetrapod) | 635 nm | ~110 meV | ~0.2 (RT) | 50% | 25 ns | - | [63] |
| PNCs | | | | | | | |
| Core only | | | | | | | |
| $CsPb(Cl/Br)_3$ | ~2.75 eV | 88 meV,<br>1 meV (6 K) | 0.045 (RT)<br>0.3 (6 K) | 32% | 0.18-0.25 ns (6 K) | 0.05 ns (6 K) | [64,65] |
| $CsPbBr_3$ | 473 nm (small) - 522 nm (large) | 72.19 meV (large) – 127 meV (small)<br>17-27 μeV (large, 3.6 K)<br>2 meV (small, 4 K) | 0.02 (small, RT) - 0.32 (large, RT)<br>0.019-0.052 (4 K) | 34.7-90%<br><br>95.6% (5K) | 0.21-0.27 ns (3.6 K)<br>0.186-0.54 ns (4 K)<br>5-11 ns | 0.0031 ns (small)-0.25 ns (large),<br>3.3 ns | [9,15,19, 26,65–75] |
| $CsPb(Br/I)_3$ | 520 nm | ~15 nm (ensemble) | 0.2 | <25% | 6.3 ns | - | [76] |
| $CsPbI_3$ | ~ 680 nm | 70 meV (large)-160 meV (small)<br>137-200 μeV (large, 4 K)<br>2 meV (small, 4 K) | 0.021-0.27 (RT), 0.54*<br>~0.05 (4 K) | 40 – 100% | 9.5 ns (small, RT)-46.9 ns (large), 55 (pure radiative)<br>0.93-1.5 (4 K) | 0.093-4.2 ns<br>~0.41 ns (4 K) | [27,65,68 ,77–83] |
| $FAPbBr_3$ | 523.5-530 nm | 87.56 meV | 0.013-0.12 (RT) | 63-91.2% | 13.8-36 ns | ~0.085-0.5 | [69,75,84 –86] |
| $Cs_{0.2}FA_{0.8}PbBr_3$ | 517.1 nm | 79.67 meV | 0.04 (RT) | 63-73% | ~15.5 ns (intensity-averaged) | - | [69] |
| Core-shell | | | | | | | |
| $CsPbBr_3/CsCaBr_3$ | 2.52 eV | > 35 meV | 0.1 (small)<br>0.16 (large) | - | ~5 ns | - | [67] |
| $CsPbBr_3/CdS$ | 531 nm | 20 nm | 0.43 | 88% | 22.8 ns | - | [87] |

**Current limitations of colloidal semiconductor quantum emitters**

Fourier-transform-limited single photons, are wave packets with a spectral bandwidth determined only by the radiative lifetime, which is difficult to achieve [88]. Indistinguishable single photons are generated when the optical coherence time ($T_2$) of QE is close to twice its radiative lifetime ($T_1$), meeting the transform limit of $T_2 = 2T_1$. The coherence time $T_2$ is related to the emitter damping rate ($\Gamma$) by the equation $T_2 = 2\pi/\Gamma$. In solid-state emitters like CQDs and PNCs, the exciton's interaction with its surrounding environment, including phonons, spin noise, and charge noise, degrades the coherence of the excited state. These interactions increase $\Gamma$ and decrease $T_2$ [26,48]. Cryogenic temperatures are often employed to reduce $\Gamma$, but even under these conditions, colloidal semiconductor QEs exhibit linewidths ranging from tens of μeV to several meV. Additionally, at low temperatures, QDs display relatively long $T_1$ values due to dark exciton states, complicating the achievement of the transform limit further. The most favorable conditions for colloidal nanocrystals have been attained with colloidal $CsPbBr_3$ PNCs and InP/ZnSe/ZnS CQDs. $CsPbBr_3$ PNCs demonstrated fast radiative recombination ($T_1 \approx 200$ ps at 4 K), and $T_2$ up to 78 ps ($\Gamma = 17$ μeV), resulting in a $T_2/2T_1$ ratio of up to ~0.19 [89],[26]. Recently, InP/ZnSe/ZnS CQDs exhibited linewidths as narrow as ~5 μeV at 4 K, yielding a lower-bounded optical coherence time $T_2$ of ~250 ps. However, the emission lifetime $T_1$ remains too long (~15 ns) to achieve transform-limited emission. Enhancing the radiative rate of these nanostructures will be key to achieving transform-limited emission.

For the generation of entangled photons, the XX→X→GS cascade recombination in colloidal nanocrystals holds potential. However, a major limitation is the low emission efficiency of the XX→X transition. The recombination of a colloidal nanocrystal from the X level to the ground state is often influenced by nonradiative recombination through internal and external defects [90]. Auger-like nonradiative recombination impacts biexciton recombination more strongly than defect-assisted relaxation [91]. The strong Coulomb interaction between charge carriers in semiconductor nanocrystal leads to high probability of energy transfer from the recombination of one electron-hole pair in a biexciton to the second electron-hol pair. Consequently, this reduces probability of radiative XX-X recombination compared to the radiative X→GS recombination. Photon-correlation spectroscopy can be utilized to measure the ratio of the radiative efficiencies of the biexciton to the exciton. In pulsed excitation scenarios, the $g^{(2)}(\tau=0)$ peak amplitude reflects the probability of detecting two photons within a single laser cycle. The ratio of the integrated area of the $g^{(2)}(\tau=0)$ center peak to the averaged side peak $g^{(2)}(\tau=T)$, denoted as $g_0^{(2)}$, is asymptotically equal to the ratio of the biexciton photoluminescence quantum yield ($QY_{XX}$) to exciton photoluminescence quantum yield ($QY_X$) when the average exciton occupancy approaches zero ($<n>\rightarrow 0$) [92]:

$$g_0^{(2)} = \frac{\int_{-\Delta t}^{\Delta t} g^{(2)}(\tau)d\tau}{\int_{T-\Delta t}^{T+\Delta t} g^{(2)}(\tau)d\tau} \approx \frac{QY_{XX}}{QY_X}, \quad (2)$$

where Δt is an appropriate integration range. In most of colloidal nanocrystals this ratio is too low to be used for the on-demand generation of photon pairs.

In addition to high emission efficiency, emitted photon pairs should exhibit a high level of entanglement fidelity for quantum information applications [29]. Ideally, they exist in the maximum entangled Bell state $|\varphi^+\rangle = (|HH\rangle + |VV\rangle)/\sqrt{2}$, with $|H\rangle$ and $|V\rangle$ being the horizontal and vertical polarization basis states, respectively. The fidelity $f_{|\varphi^+\rangle}$ of the real photon pair's state to $|\varphi^+\rangle$ is mainly determined by the fine structure splitting (FSS) $\Delta$ (Figure 3b) and lifetime $T_{1X}$ of the single-exciton state. In the absence of other dephasing mechanisms, the maximum achievable fidelity of an ensemble of photon pairs to a maximum entangled state is given by the following equation [93]:

$$f_{|\varphi^+\rangle}^{max} = \frac{1}{4} \cdot \left(1 + \alpha \cdot k + \frac{2 \cdot \beta \cdot k}{1 + (\beta \cdot \Delta \cdot T_{1X}/\hbar)^2}\right), \quad (3)$$

where $k$ is the fraction of detected photon pairs in which both photons come from the XX→X→GS cascade and not from multiphoton or background emission, which can be measured. Here $\alpha$ is the fraction of QE emission unaffected by spin scattering with a characteristic time $\tau_{SS}$, which is equivalent to the first-order cross-coherence $\beta$ in the absence of cross dephasing process with the time $\tau_{HV}$: $\alpha = \frac{1}{1+\frac{T_{1X}}{\tau_{SS}}}$; $\beta = \frac{1}{1+\frac{T_{1X}}{\tau_{SS}}+\frac{T_{1X}}{\tau_{HV}}}$. FSS in PNCs and CQDs typically ranges from 0.1-1 meV [80,94–97], and $T_{1X}$ is not sufficiently short to compensate for it. This limitation hinders the attainment of high entanglement fidelity in emitted photon pairs. Minimizing $T_{1X}$ becomes crucial to maximize entanglement fidelity. Therefore, accelerating the radiative recombination rate of the exciton state is essential to enhance the entanglement fidelity of photon pairs emitted through the XX→X→GS cascade recombination.

Low values of the quantum yield for the charged biexciton state ($QY_{XX}$) in colloidal nanocrystals also give rise to an issue related to the temporal reduction in the emission intensity of the nanocrystal, leading to PL blinking behavior [98–101]. During the XX→X transition, energy can be transferred to one of remaining charge carrier (either electron or hole) allowing them to overcome the potential barriers on the nanocrystal-environment interface and be transferred to the external trap leaving the nanocrystal in the unbalanced charge state. [99]. After subsequent absorption event a charged trion is formed [98]. In this state, PL is effectively quenched by nonradiative Auger recombination [100]. If the charged nanocrystal absorbs two photons, a charged biexciton is formed. The ratio of the PL QYs of the transitions from the charged biexciton-to-trion to the trion-to-ground state can be determined using Eq. 2. This ratio is higher than in the case of a neutral biexciton and exciton, resulting in an emission regime with very poor single-photon purity. Moreover, the spectral shift between these two transitions is lower than between the biexciton and exciton, making spectral filtering more challenging. Hence, suppressing blinking and trion formation is crucial to ensure the stability of PNCs and CQDs as single-photon emitters. Increasing the $QY_{XX}$ of colloidal nanocrystals will not affect the single-photon purity of QEs since it can be easily filtered by spectrum or lifetime [30,32].

The key insight from the above discussion is that increasing the radiative rates of the exciton and biexciton transitions is a critical task for achieving on-demand generation of both indistinguishable single photons and quantum entangled photon pairs by CQDs and PNCs. To implement this, leveraging the light-matter coupling phenomenon is necessary to increase the radiative rate by means of the Purcell effect. Both optical cavities and plasmonic antennas and nanocavities can be utilized for this purpose. However, the maximum possible values of the Purcell factor are achievable by using plasmonic nanocavities due to their smaller electromagnetic mode volumes [102] (Figure 4).

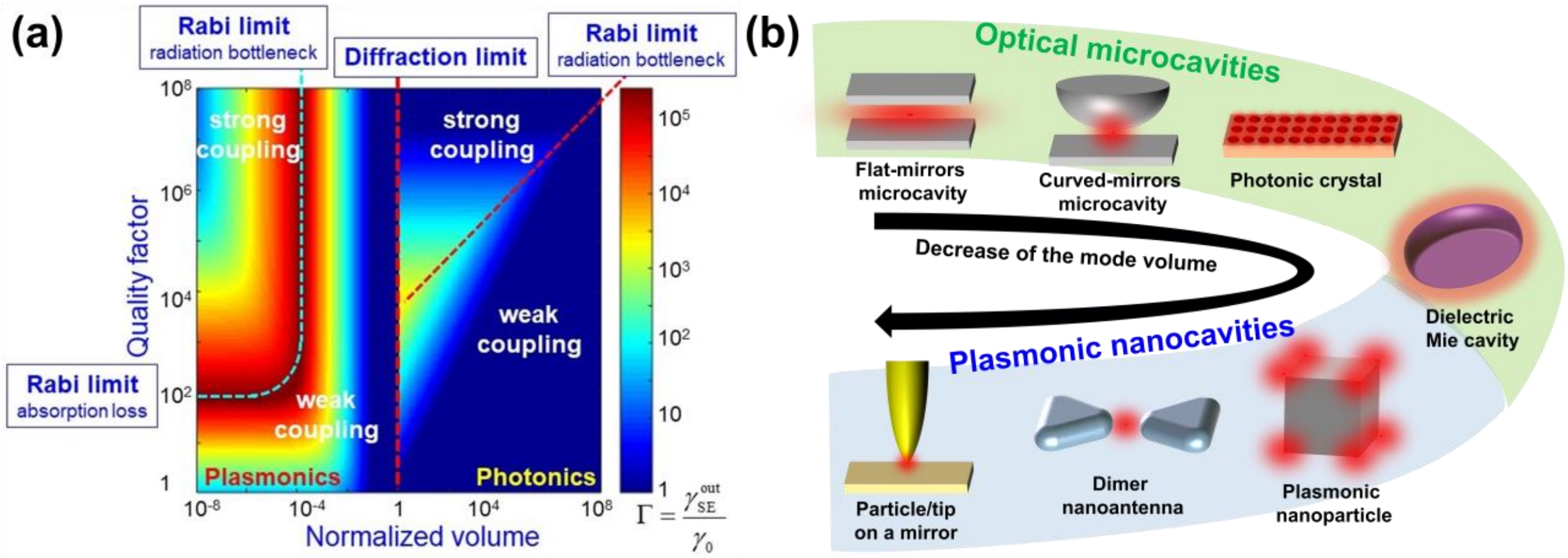

**Figure 4.** Comparison of photonic microcavities and plasmonic nanocavities for the maximization of the Purcell factor and exciton radiative rate. (a) The enhancement of the out-of-cavity emission rate as a function of the normalized volume and quality factor for both plasmonic (lossy) and photonic (diffraction-limited) configurations. Figure obtained from a private conversation with authors of the Ref. [102]. (b) Scheme representing different cavity types in order of reducing the cavity mode volume.

## 2. Plasmon-exciton quantum emitters based on colloidal nanocrystals

### 2.1 Light-matter coupling phenomenon

Light-matter coupling is a phenomenon that commonly occurs when the energy transitions of excitons, molecules, or atoms (quantum oscillators) are in resonance with the electromagnetic mode of a cavity they are placed in (Figure 5a,b) [39,45,103–107]. This phenomenon manifests itself in various ways depending on the cavity parameters and the type of the transition and its oscillator strength. The main parameter for the light-matter coupling phenomenon is a coupling strength (g), which can be calculated using the following equation:

$$g = \sqrt{N}\mu_{ij}\sqrt{\frac{h\omega_0}{2\varepsilon\varepsilon_0 V_{cav}}}, (4)$$

where $N$ is the number of equal oscillators involved with the interaction with the selected cavity mode, $\mu_{ij}$ is the dipole moment of the oscillator transition, $\omega_0$ is a resonance frequency (equal for the exciton transition and cavity mode), and $V_{cav}$ is the cavity mode volume. For all single QE applications, $N = 1$. Depending on the ratio between the coupling strength and damping rates of the cavity mode ($\gamma_{cav}$) and the transition ($\gamma_{ex}$), two main regimes of light-matter coupling are possible: weak light-matter coupling (Figure 5c) and strong light-matter coupling (Figure 5d). The main condition for the weak coupling regime is $< \gamma_{ex}, \gamma_{cav}$. In most semiconductor QEs, there are many possible excitation or absorption transitions, while the emission transition commonly occurs from the lowest-energy exciton level [108,109]. Thus, for semiconductor QEs, it is necessary to distinguish light-matter coupling involvement of absorption transition and emission transitions. Weak coupling may occur between the cavity mode and the absorption transition, manifesting itself as exciton absorption enhancement and the induced transparency phenomenon of the whole system [107]. For QE applications, exciton absorption enhancement is most crucial as it leads to more efficient excitation of QEs and, consequently, higher emission brightness. In the case of weak coupling of the exciton emission transition with a cavity, the Purcell effect is implemented. The Purcell effect is the result of the increase in the local density of optical states inside the cavity mode, leading to an increase in the emission rate of the exciton. The ratio of the emission rate of the exciton coupled to the plasmon ($\Gamma_{cav}$) to the initial exciton emission rate ($\Gamma_{free}$), known as Purcell factor, $F_p$, can be estimated by following equation:

$$F_p = \frac{\Gamma_{cav}}{\Gamma_{free}} = \frac{3}{4\pi^2}\left(\frac{\lambda}{n}\right)^3 \frac{Q}{V_{cav}}, (5)$$

where $\lambda$ is the emission wavelength (in vacuum), n is the refractive index of the cavity material, $Q = \frac{\omega_{cav}}{\gamma_{cav}}$ is the quality factor of the cavity, and $V_{cav}$ is the cavity mode volume. Significant emission rate enhancement requires an optical resonator that confines light to small dimensions (small $V_{cav}$) and that stores energy for a long time (high $Q$). While quality factors of photonic cavities are usually very high (~$10^4$ for photonic crystals), the minimum possible mode volume is limited by the diffraction limit. [103]. In the case of using plasmon cavities, electromagnetic plasmon oscillations play the role of the light in the coupling, and the corresponding modes can be below the diffraction limit, down to several nanometers inside of nanogaps. Thus, using plasmonic nanocavities enables to achieve higher values of the Purcell factor than using photonic cavities (Figure 4) [102]. However, plasmonic cavities are characterized by strong damping due to optical losses in the metals, which results in general in low quality $Q$ of the plasmonic cavities.

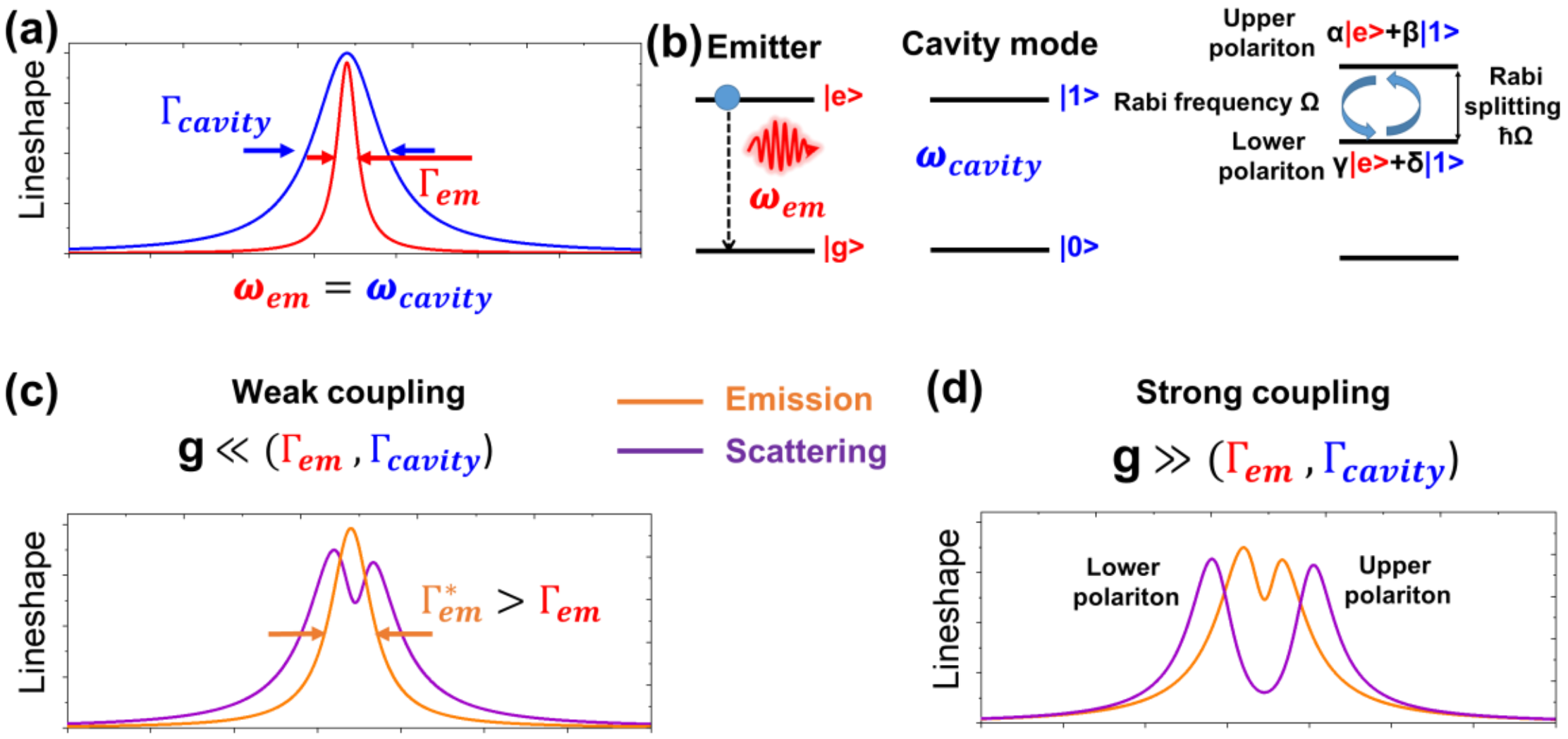

**Figure 5.** The principles of light-matter coupling of the emitter with a cavity. (a) Conditions for the light-matter coupling; (b) polaritons formation; (c) optical properties of weakly coupled system; (d) optical properties of strongly coupled system.

The other light-matter coupling regime is strong coupling, with the main condition being $g > \gamma_{ex}, \gamma_{cav}$ [103]. However, the lower limit for $g$ ensuring the strong coupling regime is not the same in different studies. The less strict criterion for the coupling strength is $2g > (\gamma_{cav} + \gamma_{ex})/2$ [110] and the most strict is $g > \sqrt{({\gamma_{cav}}^2 + {\gamma_{ex}}^2)/2}$ [111]. One of the main results of strong coupling is the splitting of the exciton transition energy level into two polaritonic levels, combining properties of both light and matter. The energy separation between these levels is called Rabi splitting, and its energy ($\Omega_R$) is given by $\Omega_R = 2\sqrt{g^2 - (\gamma_{cav} - \gamma_{ex})^2/4}$ [111,112] also known as the Rabi energy or Rabi frequency. This parameter can be measured experimentally. For practical use, the strong coupling criteria could be applied to the Rabi energy with the less strict condition $\Omega_R > |\gamma_{cav} - \gamma_{ex}|/2$ [113] and the most strict condition $\Omega_R > \gamma_{cav} + \gamma_{ex}$ [111]. The effect of strong coupling on the emission properties of QEs could be much more complex than just the Purcell effect due to the change in the quantum properties of the state from which the emission is realized [39,103–106].

## 2.2 Plasmon-exciton quantum emitters in weak coupling regime

### *2.2.1 Emission intensity enhancement and lifetime shortening*

In the realm of weak plasmon-exciton coupling, three distinct phenomena come to the forefront — induced transparency, exciton absorption enhancement, and the Purcell effect. However, concerning QEs, it becomes evident that only the Purcell effect and exciton absorption enhancement significantly impact the emission properties. In Table 2, we have compiled a comprehensive summary of studies showcasing the demonstration of plasmon-exciton coupling for confirmed individual QEs, evidenced by the measurement of the second-order correlation function, highlighting the most important parameters. For more examples, including 2D materials, refer to Tables S2A and S2B in the Supplementary material file.

**Table 2.** Emission properties of plasmon-exciton QEs.

| Cavity geometry | Plasmon res., nm | Quantum emitter | PL max, nm | Uncoupled/Coupled | | PL increase factor | Purcell factor | Ref. |
|---|---|---|---|---|---|---|---|---|
| | | | | Lifetime, ns | $g^{(2)}(0)$ | | | |
| **Bowtie nanoantennas** | | | | | | | | |
| Deformed Au bowtie | 800 | CdSe/ZnS QD | 655 | 35 / 19 | 0.41 / 0.38 | 7 | - | [114] |
| Parallel Au nanobars | 660 | CdSe/CdS QD | 640 | 40-80 / 5-45 | ~0.15 / 0.3-0.7 | - | 1.4-9.6 | [115] |
| Au nanorod dimer | ~620 | CdSe/ZnS QD | 620 | 14.2 / 5 | 0.25 / 0.5-0.65 | 1.6-2.2 | - | [116] |
| Ag bowtie | 636 | CdSe/ZnS QD | 633 | 22 ns / ~6-7 ns | 0.25-0.4 / 0.5-0.8 | - | - | [117] |
| **Nanopatch antennas** | | | | | | | | |
| Ag nanocube / Au film | 630 | CdSe/ZnS QD | 630 | 6.8 / 0.013 | 0.17 / 0.32-0.46 | 1900 | 540 | [118] |
| Ag nanocube / Au film | 560 | CdZe/ZnS QD | 655 | 25 / 1 | 0.1 (averaged) / 0.7-1.0 | ~ 4 | ~70 | [119] |
| Au nanodisk / Au film | - | CdSe/CdS QD | 635 | 36 ns / < 1.5 ns | 0.2-0.3 / 1 | 72 | 72 | [120] |
| **Bullseye antennas** | | | | | | | | |
| Au bullseye | 617 | CdSe/CdS QD | 2.005 eV | 1.63 / <0.23 | - / <0.5 (~0.4) | 5.5 | 7 | [121] |
| | | | 2.004 eV | 60.5 / 2.2 | | | 28 | |
| Ag bullseye | - | CdTe/ZnS QD | 780 | 170 / - | 0.12 / 0.37 | - | - | [122] |
| Au bullseye | 660-680 | CdSe/CdS QD | ~660 | 67 / - | - / 0.68 (0.07 filtered) | - | - | [123] |
| **Single nanoantenna** | | | | | | | | |
| Ag covered AFM tip | 450 | CdSe/ZnS QD | 610 | 15.125-26.7 / 0.3-0.4 (AWL) | 0.08-0.09 / 0.85- 0.88 | < 2.5 | - | [124] |
| Au nanocone | 625 | CdSe/CdS QD | 650 | X: 62 / 1.6 XX: 4 / 0.5 | 0.30 / 1.18 | ≤ 45 | X: ~109 XX: ~100 | [125] |
| Au nanocube | 600 | CdSe/ZnS QD | 610 | 29.6 / 0.4 | 0.13-0.14 / 0.97 | 1.3 | 103 | [126] |
| Au nanocube | 680 | CdSe/CdS/ZnS QD | 620 | 17.5 / 0.63 | 0.2-0.35 / 0.35 | - | 24 | [127] |
| Spherical Ag NP@$SiO_2$ | 439-446 | CdSe/ZnS QD | 605 | 7.72 / 0.56 (AWL) | 0.09 / 0.81 | ~ 2 | - | [128] |
| Ag nanoprism | 700 | CdSe/ZnS QD | 600 | 25 / 5 | - / 0.2 | 2.5 | 12.5 | [129] |
| Ag nanowire | - | CdSe/ZnS QD | 650 | 25.4 / 3.2 | - / 0.3 | - | 17.3 | [130] |
| Ag nanowire | - | CdSe/ZnS QD | 650 | 24.2 / 5.6 | - / 0.29 | - | - | [131] |
| Ag nanowire | - | CdSe/ZnS QD | 655 | 22 / 13 | - / <0.3 | - | ~1.6-2.5 | [132] |
| **Ordered nanoantenna arrays / metasurfaces** | | | | | | | | |
| Ag PNP array | 582 | CdSe/ZnS QD | 570 | 29 / 6.8 (AWL) | 0.02 / 0.34 | 1.5 | 6 | [133] |
| Ag PNP array + PDMS | 510 | | | 24 / 1.67 (AWL) | 0.018 / 0.27 | 4.2 | 22 | |
| Ag 1D grating | - | CdSe/ZnS QD | 600 | 17.1 / 0.7-7.3 | 0.06 / 0.21-0.25 | 3-5.3 | 3.6 | [134] |
| **Disordered nanoantenna arrays / metasurfaces** | | | | | | | | |
| Silver nanocube metasurface | 515 | $CsPbBr_{2.5}I_{0.5}$ PNC | 520 | 0.8-1.1 / 0.4 | <0.5 / >0.5 | 2.2-7.5 | 5-12 | [76] |
| | | CdSe/CdS QD | 620 | 5.8 / 0.6 | <0.5 / <0.5 | ~ 0.24 (decrease) | - | |
| | | Core/multishell CdSe/ZnS/CdS/ZnS QD | 560 | 3.2 / 0.5 | <0.5 / >0.5 | ~ 1.3 | - | |
| Ag nanoprisms | 570 | Core/multishell CdSe/ZnS/CdS/ZnS QD | 560 | 3.7 ns / 0.3 – 0.5 | 0.8-1.3 | 0.1-1.9 | 23-37 | [44] |
| Ag PNPs | 430 | Core/multishell CdSe/ZnS/CdS/ZnS QD | 560 | 3.7 ns / 0.4 ns | 0.9 – 1.3 | ~ 1.5 (average) | - | |
| Au nanorods | 610 | Core/multishell CdSe/ZnS/CdS/ZnS QD | 560 nm | 18 ns / 1.5 ns | 0.1-0.3 / 0.712-0.968 | ~ 0.33 (decrease) | ~3 | [43] |
| Spherical Au NPs@$SiO_2$ | 612 | CdSe/CdS QD | 629 | 37.93 / 4.03-22.67 (AWL) | 0.1-0.2 / 0.6-0.8 | - | X: 1.4 XX: 5.6 | [135] |
| Ag NPs | 430 | CdSe/ZnS QD | 610 | 23 / 0.5 | 0.05-0.2 / 0.05-0.5 | 2.5-4 | - | [136] |
| Ag nanoflakes film | - | CdSe/CdS QD | 620 nm | 39 / 0.9 | 0.1 / 0.9-2 | 0.5 (decrease) | 4 | [137] |
| Ag NPs | 590 | CdSe/CdS QD | 638 nm | - | 0.021-0.045/1.16 | 2.5-11.9 | - | [138] |
| Au NPs | 530 | CdZnSe/ZnS | 532 | 15 / 2 | 0.13 / 0.69 | 2-6 | 2-5 | [139] |
| **Au films** | | | | | | | | |
| Au film | - | CdSe/ZnS QD | 617 | ~18 / 6-8 | 0 / ≤0.45 | - | X: 1.5 XX: >10 | [140] |
| Au film | ~650 | CdSe/CdS QD | 660 nm | 75 / 0.8-2.9 | ~0 / ~1 | 2-7 | 14-60 | [141] |
| Au film | ~650 | CdSe/CdS QD | 620 nm | 60 / 6-40 | - / <0.2 | 1.5-10 | 10 | [142] |
| **QE-in-shell** | | | | | | | | |
| Au shell | 750 | CdSe/CdS/$SiO_2$ QD | 670 | 123 / 20 | 0.2-0.28 / 0.74-0.90 | 1.12 | ~6 | [143] |

The Purcell effect in plasmon-exciton systems can lead to both enhancements and reductions in emission intensity. In classical systems, where emitters are coupled to optical microcavities, the cavity typically exhibit a lower damping rate than that of the emitter, with nonradiative losses determined solely by emitter properties. The Purcell factor in such systems denotes the increase in the pure radiative rate of the emitter transition. However, in plasmonic systems, nonradiative losses are significantly higher than in optical cavities, leading to an increase in both radiative and nonradiative recombination of excitons coupled to them. Consequently, two key definitions emerge for the plasmon-exciton Purcell effect—the classical definition, considering only radiative rate enhancement, and the broader interpretation, encompassing the calculation of total emission rate enhancement. When the emission transition of QE is coupled to a plasmon mode, the energy of the exciton can be transferred to the plasmon mode, subsequently being emitted into the far field or dissipated as heat. This phenomenon was experimentally validated by Akimov et al. [132] through the coupling of a single QD to plasmons in a 3 µm long silver nanowire. The signal transferred to the plasmons was detectable from the end of the nanowire, distinct from the photons emitted solely from the location of the QD. Photon correlation between the CQD emission and the ends of the nanowire confirmed the generation of single, quantized plasmons stimulated by the energy transfer from the QD. A similar effect was observed in the study conducted by Li et al. [131]. The efficiency of plasmon mode scattering, depending on the specific plasmon mode to which the exciton emission transition is coupled, can lead to a significant increase [120,125,134] or even a decrease [43,44,76,144] in the total PL QY of colloidal single QE. Furthermore, an additional metal-induced quenching process may occur due to charge carrier transfer [137], further reducing the emission efficiency.

To enhance emission intensity, utilizing the effect of the plasmon-induced exciton absorption (or excitation) enhancement can be more efficient than the Purcell effect. The origin of exciton absorption enhancement can be attributed to the substantially higher absorption cross-section of plasmon structures in comparison to CQDs and PNCs [145]. This difference allows plasmon structures to store considerably more energy than a single exciton or even a multiexciton state within a single QE. Particularly, if the plasmon mode resonates with the absorptive exciton transition of the QE, energy transfer from the plasmonic structure to the QE becomes probable. In case where the photon absorption of a single QE from an external light source is not in the saturation regime, the likelihood of energy transfer from the plasmonic structure significantly increases. This scenario leads to an effective plexciton absorption cross-section surpassing the exciton absorption cross-section itself [146], resulting in the augmentation of the emission intensity.

In general, one can distinguish 8 different geometries of plasmon nanoantenna-single colloidal QE systems presented so far in the literature (Table 2, Figure 6). These include: (1) nanogap antenna, where the QE is placed in a nanometer-sized gap between two plasmonic antennas; (2) nanopatch (particle-on-the-mirror) antenna, composed of single plasmonic particle on top of metal film acting as a mirror with QE in between; (3) single nanoantenna, usually an individual plasmonic nanoparticle or metal-covered AFM tip placed in proximity to the QE; (4) ordered and (5) disordered arrays of nanoantennas, including plasmonic metasurfaces, where plasmon resonance is a collective response of multiple antennas in an array; (6) simple metal films; (7) bullseye antennas, which may additionally increase the directionality of emission and (8) plasmon shell on the surface of QD. The most significant enhancement in emission was achieved by Hoang et al. [118]. They utilized the nanopatch antenna geometry (particle-on-the-mirror) with a silver nanocube on a gold mirror, resulting in a robust confinement of the electromagnetic mode. This configuration led to a remarkable 540-fold increase in the spontaneous emission rate and approximately a 1900-fold enhancement in time-averaged emission intensity. The PL QY of the emitter, modified by the nanocavity, is predicted to be 50%, a substantial improvement compared to the estimated 20% QY for QDs alone. The primary contributor to the emission enhancement was the 225-fold increase in the excitation rate. The remainder of the signal enhancement was attributed to the enhanced photon collection efficiency, reaching up to 84%, collected by a 0.9 NA objective due to the highly directional photon emission from the plexciton QE. It is noteworthy that CW excitation was employed, raising the possibility of faster QD relaxation and subsequent re-excitation. In contrast, an earlier work by Yuan et al. [119] achieved only a fourfold emission enhancement (with a Purcell factor of 70). The key distinction between the studies of Hoang et al. and Yuan et al. lies in the perfect overlap of QD emission with the scattering mode of the plasmonic system in the former, while the latter did not achieve resonance. Consequently, the weak coupling efficiency of QD emission to the scattering plasmon mode in Yuan et al. prevented the emission of photons transferred to the plasmon mode from the exciton. The second most significant emission enhancement, a factor of 72 (see Table 2), was reported by Dhawan et al. [120]. They employed the nanopatch geometry with a gold nanodisk and gold mirror.

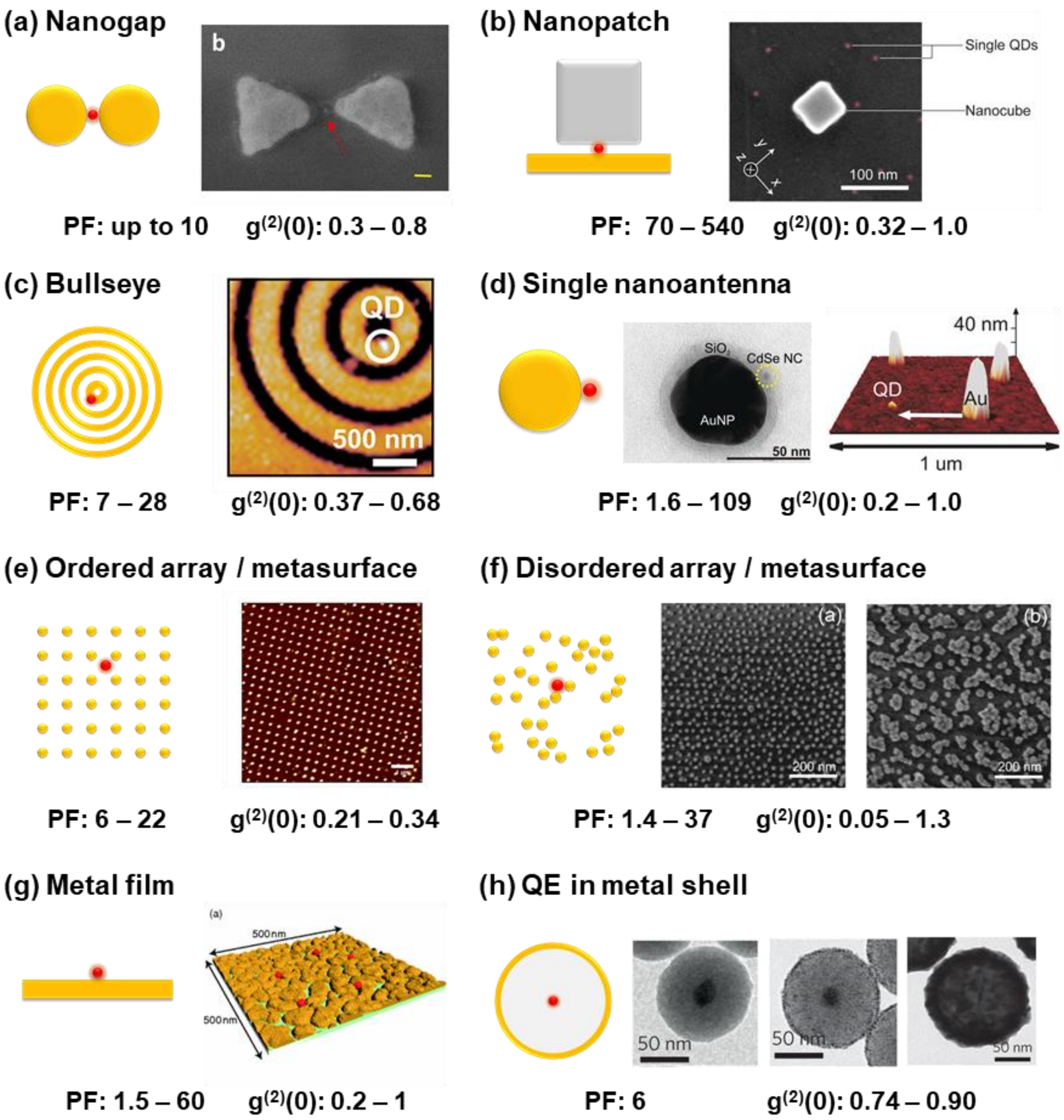

**Figure 6.** Different geometries of plasmonic nanoantenna-QE systems. For each geometries the ranges of Purcell factor (PF) and $g^{(2)}(0)$ for a coupled systems obtained in a literature so far are given. (a) STEM image of Ag bowtie nanoantenna with QD in the gap (red arrow). Scalebar is 20 nm. (b) SEM image of a Ag nanocube on top of Au film with individual QDs. (c) AFM micrograph of a portion of the bullseye antenna with QD located on an inner circle. (d) Left: Single Au nanoparticle with $SiO_2$ shell and individual QD attached. Right: AFM image of a single QD and Au NPs with a white arrow denoting the path of the Au NP displacement. (e) AFM image of the square lattice of Ag nanoparticles. Scalebar is 1 µm. (f) SEM image of the disordered array of Au NPs on a glass. Left: individual Au NPs. Right: Islands of aggregated Au NPs. (g) AFM topography of the rough gold film. (h) TEM images from different stages of preparation of a Au shell. Left: single QD with thick $SiO_2$ shell. Middle: QD@$SiO_2$ with Au seeds on the surface. Right: QD@$SiO_2$ covered with 20 nm thick Au shell. (a) Adapted from [116]. (b) Adapted from [118]. (c) Adapted with permission from [121]. Copyright 2024 American Chemical Society. (d) Left: Adapted with permission from [147]. Right: Adapted with permission from [148]. Copyright 2024 American Chemical Society. (e) Adapted with permission from [133]. Copyright 2024 American Chemical Society. (f) Adapted with permission from [149]. (g) Reprinted figure with permission from [141]. Copyright 2024 by the American Physical Society. (h) Adapted from [143].

High photon collection efficiency can also be attained using bullseye nanoantennas. Livneh et al. [122] demonstrated that more than 35% of photons emitted from a single QD are collected in a silver bullseye nanoantenna with NA = 0.65 objective. Abudayyeh et al. [123] achieved a record-high 85% collection efficiency of emitted photons into a low NA = 0.5 using a gold bullseye nanoantenna. This type of plasmonic nanostructures also resulted in an emission enhancement of up to 5.5 with a Purcell factor reaching 28 for single QDs [121].

While nanogap antennas provide strong confinement of the electromagnetic mode and are suitable for achieving strong plasmon-exciton coupling modes [116], experimentally achievable Purcell factors do not exceed 10 [115], and the maximum achieved PL enhancement is only by a factor of 7 [114].

The simplest geometry of a single plasmonic nanoparticle can yield a wide range of Purcell factors and emission enhancement factors depending on the particle geometry and the QE's position relative to the hot spots of resonant scattering plasmon modes. For instance, the Purcell factor varies from 1.6 in the case of a QD positioned at the side of a silver nanowire [132] to 109 in the case of the positioning of a QD directly at the end of a gold nanocone, resulting in approximately 45 times emission enhancement [125]. In the geometry of ordered plasmonic arrays, emission enhancement with a factor up to 5.3 [134] and a Purcell factor up to 22 [133] can be achieved.

The utilization of thin metal films allows QDs interact with surface plasmon polaritons instead of localized plasmons. The highest Purcell factor was achieved for rough nanostructured gold films with pronounced plasmon resonance [141,142], as opposed to flat and homogeneous metal films [140]. Similar characteristics are observable for disordered plasmonic arrays and metasurfaces. In contrast to ordered plasmonic arrays, the random distribution of QDs still allows for good coincidence with the hotspots of the metasurface due to the high density of plasmonic structures. This enabled the measurement of the same single QDs before and after the deposition of a plasmonic array (metasurface) in the works of Krivenkov et al. [43,44] and single PNCs and QDs in the work of Olejniczak et al. [76], with a maximum achieved Purcell factor of 37 [44] and emission enhancement of up to 7.5 [76]. Furthermore, these studies experimentally demonstrated the importance of strong spectral overlap between exciton emission transitions and plasmon modes for Purcell-induced emission enhancement in QDs and PNCs interacting with metasurfaces.

The metal shell on the surface of a single QD [143] represents a rare geometry, exhibiting only a 12% enhancement of QD neutral state emission with a Purcell factor of 6. The lack of strong field localization at the QD position in this system may be a cause for this limited enhancement. At the same time, the strong blinking suppression is the important result of the study. In general, blinking suppression can be a result of the recombination rate enhancement [43,44,141], and increase of the emission QY [117,119,143]. This blinking suppression leads to the increase in the average emission intensity. At the same time, the quenching of the on-state exciton emission also leads to the blinking suppression accompanying with intensity reduction [128]. Masuo et al. demonstrated that off-states in the QD fluorescence can be effectively suppressed by the combination of the biexciton emission enhancement and the quenching of the exciton one [126].

An additional contribution to the total emission enhancement can be made through enhanced biexciton emission due to the XX→X transition (Figure 3). The resonant excitation of the biexciton state is a nonlinear two-photon process. Thus, a significant increase in photon absorption efficiency due to plasmon coupling could markedly influence efficient biexciton generation. Additionally, the biexciton PL QY could be substantially enhanced by the Purcell effect, altering the QE photon statistics and transforming semiconductor colloidal QEs into sources of photon pairs instead of single photons. However, the change in the biexciton emission efficiency can alter the photon statistics of semiconductor colloidal QEs.

*2.2.2 Photon statistics altering*

One crucial parameter defining the quality of QEs is the statistics of emitted photons. In the context of generating single, indistinguishable photons, the ideal scenario involves achieving a $g_0^{(2)}$ value lower than 0.5, approaching 0, indication near-perfect single-photon emission. However, the interaction of single colloidal QEs with plasmonic metal nanostructures introduces a significant complication—shifting the emission statistics from single-photon antibunching ($g_0^{(2)}$< 0.5) to multiphoton emission ($g_0^{(2)}$> 0.5, refer to Table 2). This phenomenon was convincingly illustrated in studies by Takata et al. [124], Matsuzaki et al. [125], and Masuo et al. [126], which investigated the coupling of CQDs to plasmonic structures. Employing AFM-based technique, they precisely manipulated the distance between the QDs and plasmon structures, either by moving Au nanoparticles on the surface towards QDs or by direct approaching of Ag-covered AFM tip to single QD, inducing the plasmon-exciton interaction. A significant advantage of this approach is its reversibility, allowing observation of changes in $g_0^{(2)}$ and lifetime before, during and after the coupling process (Figure 7a-d). Subsequent research by Krivenkov et al. [43,44] extended this investigation by measuring the same CQDs before and after the deposition of plasmonic metasurfaces. This revealed alterations accompanied by a notable shortened lifetime, blinking suppression [43,44], and emission enhancement [44] at relatively long plasmon-exciton distances up to 100 nm. In the case of PNCs, a significant breakthrough was achieved in the work of Olejniczak et al. [76]. They demonstrated the change of emission statistics from single-photon to multiphoton, with a pronounced Purcell effect. Using a plasmonic metasurface consisting of silver nanocubes on a curved glass substrate (Figure 7e,f), they showcased the reversible nature of this transition. Importantly, this technique was proven effective not only for single PNCs and CQDs but also for arrays and ensembles of QEs when their emission spectrum resonates with the plasmon mode of the metasurface. This stands in contrast to AFM-based techniques, emphasizing the broader applicability and versatility of metasurface-mediated control of emission statistics.

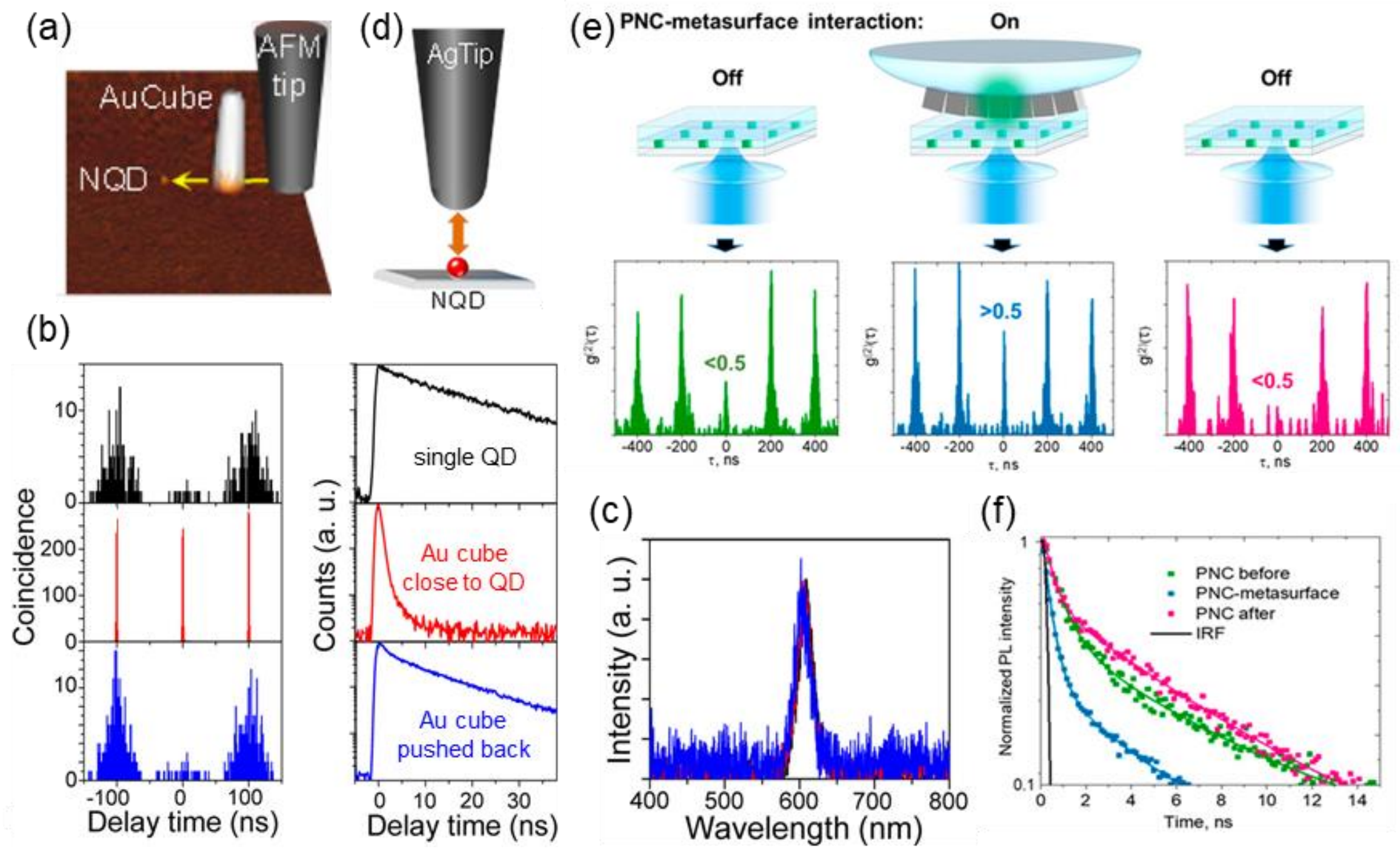

**Figure 7.** Examples of reversible weak coupling of individual single photon emitters to plasmonic nanoantennas. (a) AFM image with single QD and Au nanocube on a glass surface. Au nanocube is pushed by AFM tip in the direction indicated by yellow arrow. (b) Second-order correlation function (left panel) and PL decay curves (right panel) measured from a single QD before (black), during (red) and after (blue) attaching an Au nanocube to it. (c) PL spectra of individual QD as described in (b). (d) Reversible coupling with results similar to presented in (a-c) was obtained also by approaching Ag-covered AFM tip vertically to the single QD on a glass. (e) Second-order correlation function measured from single perovskite nanocrystal before (green), during (blue) and after (pink) interaction with Ag nanocube metasurface. (f) PL decay curves of an individual perovskite nanocrystal as described in (e). (a, b, c) Reprinted with permission from [126]. Copyright 2024 American Chemical Society. (d) Reprinted with permission from [124]. Copyright 2024 American Chemical Society. (e, f) Reprinted from [76].

In general, in the plasmon-exciton structures with good resonant conditions between exciton emission and plasmon band the most of studies show the change of the emission statistics (Table 2). The most popular explanation of the nonzero value of $g_0^{(2)}$ in these studies was the enhancement of biexciton (or multiexciton in general) emission comparing to the single-exciton emission. As mentioned earlier, the coupling of exciton transitions to scattering plasmon modes can lead to the increase of the PL QY of excitons. Due to the relatively low energy difference between single-exciton transitions and biexciton transitions (Figure 2c and 2f) and simultaneously low-quality factor of plasmon modes, selective coupling between the single-exciton or biexciton transition remains a challenging goal that has not yet been achieved. Thus, increasing the single-exciton PL QY is also accompanied by an increase in the biexciton PL QY, leading to the rise of the $g_0^{(2)}$ [43,76,115,117,119,120,125,143]. Excitation enhancement can also lead to the stronger enhancement of biexciton emission due to the nonlinear two-photon character of biexciton excitation[123].

However, changes in photon statistics can be caused by other factors. Stronger quenching of exciton emission compared to biexciton emission can lead to  change in emission statistics towards multiphoton emission [44,124,128,137]. According to the equation (2), $g_0^{(2)}$ is equal to the ratio of the $QY_{XX}$ to the $QY_X$. Thus, a greater reduction in $QY_X$compared to $QY_{XX}$ will result in an increase in the $g_0^{(2)}$ value. Park et al. [137] and Krivenkov et al. [44] demonstrated that a reduction in $QY_X$due to the interaction with the surface of metal nanostructures [137] and plasmon nanoparticles [44] can lead to a significant increase in $g_0^{(2)}$, even above 1, resulting in super-poissonian emission statistics. Masuo et al. considered both the enhancement of biexciton emission rate and the quenching of exciton emission as  reasons for multiphoton emission mode [126]. Another possible reason for super-poissoninan emission statistics is the strong plasmon-induced increase in excitation efficiency, leading to an increase in the average exciton occupancy of single QD. Dey et al. demonstrated this by varying the overlap between the excitation wavelength and plasmon mode [138]. In their previous work, Dey et al. [135] considered all of aforementioned effects responsible for changes in photon statistics.

In the work of Livneh et al. [122] an alternative explanation of the multiphoton emission statistics from plasmon-exciton coupled QDs was proposed. The authors found that a silver bullseye nanoantenna can produce a signal in the absence of QDs. Moreover, they showed that during the off-periods of QD emission, the emission statistics is multiphoton, while during the on-periods, it is single-photon, attributing this to the background emission from the silver substrate. However, this explanation of the change in multiphoton emission can't be applied to all plasmonic nanostructures. In the study by Olejniczak et al. [76], the same plasmonic metasurface made of silver nanocubes was used for the interaction with two types of CdSe-based CQDs, one of which had an emission band in the resonance with plasmon band, and the second one did not. While in the case of the absence of resonance, the emission statistics was single-photon, reaching resonant conditions led to a change towards multiphoton emission. The excitation conditions were similar for both QD types. Thus, the observed effect of the $g_0^{(2)}$ increase could not be attributed to the background emission of silver nanoparticles. Additionally, in the study by Masuo et al. [126], the emission spectrum of QD remained unchanged before and after coupling to the gold nanocube, despite the change in photon statistics from single-photon to multiphoton.

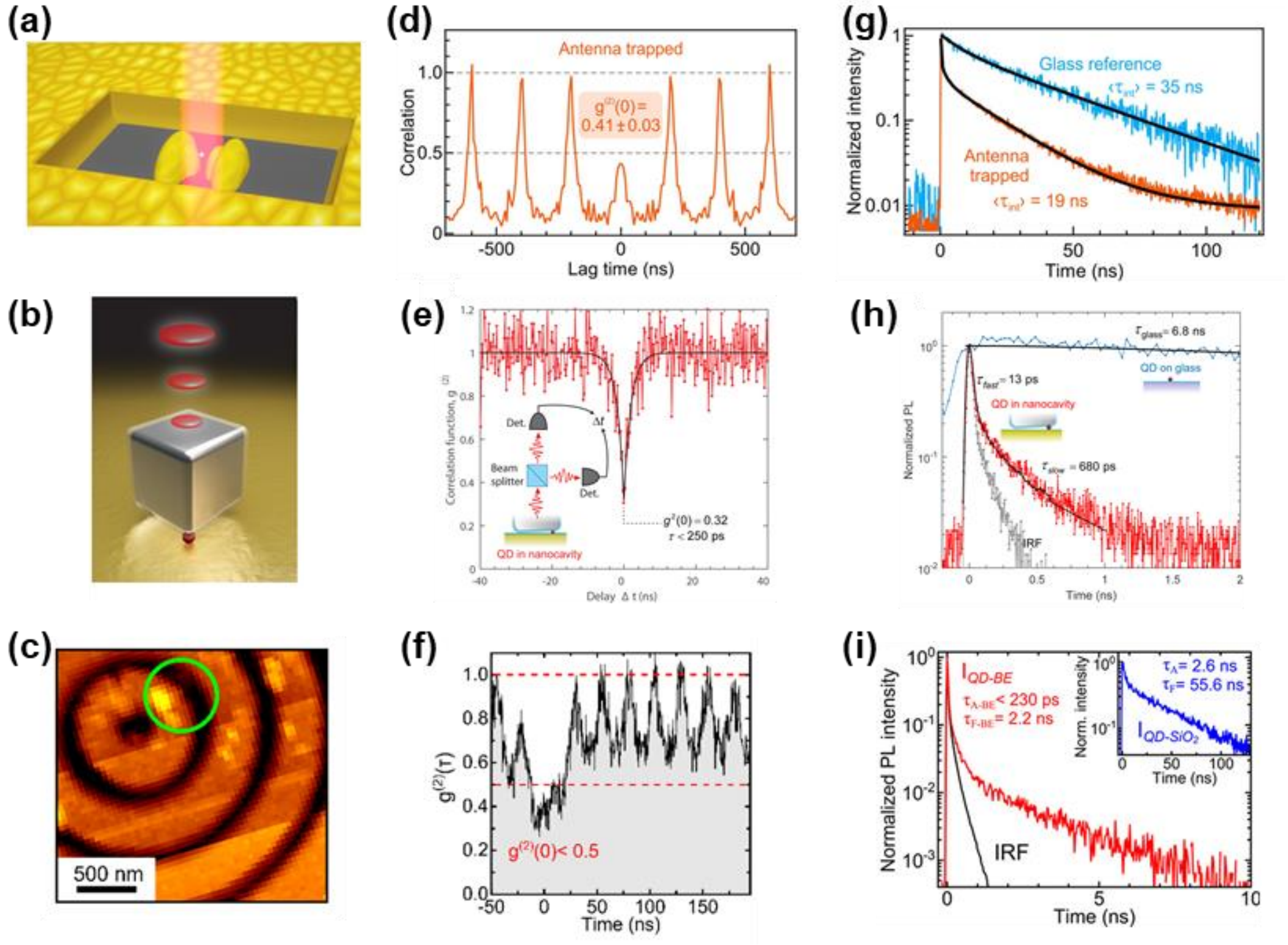

**Figure 8.** Single CQDs coupled to plasmonic nanocavities maintaining single photon emission. (a, b, c) Various geometries of nanocavities: (a) Schematic representation of nanogap Au antenna with QD placed inside a gap with 1064 nm laser trapping. (b) Illustration of Au-film/QD/Ag-nanocube nanopatch antenna. (c) AFM micrograph of Au bullseye antenna with QD positioned close to its center (green circle). (d, e, f) Second-order photon correlation measurements from a nanocavities with single QD presented in (a,b,c) respectively. Values of $^{(2)}$ (0) < 0.5 indicate single photon emission mode. (g, h, i) PL decay curves for single QDs in nanocavities presented in (a, b, c) respectively (orange and red lines) compared with reference singe QDs on glass (blue lines). (a, d, g) Reprinted with permission from [114]. Copyright 2024 American Chemical Society. (b, e, h) Reprinted from [118]. (c, f, i) Reprinted with permission from [121]. Copyright 2024 American Chemical Society.

To achieve single-photon emission statistics from CQEs coupled to plasmon nanostructures, careful design of the system is needed. In contrast to previous scenario, stronger quenching of exciton emission compared to the biexciton one should be avoided, as well as a stronger increase in $QY_{XX}$ compared to $QY_X$. Additionally, the excitation efficiency should be sufficiently low to prevent multiple exciton occupation of nanocrystals. In studies by Li et al. and Akimov et al [131,132] authors have shown that single QDs coupled to nanowires with unknown plasmon spectrum exhibit single-photon statistics. In these cases, no significant background emission of metal nanowires themselves was observed. However, the shortening of lifetime was minimal in both cases, less than 4 times, and no significant emission enhancement was observed. Therefore, in both studies, no enhancement of QY was expected. In most studies, achieving single-photon emission was enabled by poor resonant conditions between the main scattering plasmon mode and exciton emission [129,133] or the absence of this resonance [76,114,127,130,134] (Figure 8a,d,g).

In some studies involving resonant conditions, an increase of $g_0^{(2)}$ was observed. However, if its level remains below 0.5, the emission can be considered single-photon (Figure 8b-i). In the investigations by Hoang et al. [118] and Werschler et al. [121], the QD emission was in the resonance with the plasmon structures, resulting in $g_0^{(2)}$ increase to 0.46 [118] and almost 0.5 [121], respectively. It should be noted that even under non-ideal resonant conditions, $g_0^{(2)}$ can increase compared to uncoupled QDs. In the study by Yadav et al. [133], coupling a single QD with localized or collective plasmon modes of an array of silver nanostructures led to $g_0^{(2)}$ increasing from 0.02 to 0.27 and 0.34, respectively. Mallek-Zouari et al. [142] observed single-photon emission under weak resonance conditions between QD emission and plasmon mode, in contrast to their previous study [141], where the same plasmonic structure, but with better resonance conditions, resulted in multiphoton statistics. Additionally, it also is important to note that achieving single-photon emission required the use of a 30 nm thick silica spacer between the gold film and the QD, unlike the study where multiphoton emission was achieved without a spacer [141]. Thus, both the absence of quenching and weaker plasmon-exciton coupling conditions (including the Purcell effect) could impact the single-photon purity of the emitting signal.

The post-processing technique of temporal filtering could be employed to differentiate exciton and biexciton emission [7]. In the study by Abudayyeh et al. [123] gold bullseye nanoantenna was utilized for directive and enhanced emission of single QD. However, measurements were primarily conducted near saturation, where both biexciton and plasmonic emission play a role. Given that the emission of single excitons is at least 4 times slower than that of biexcitons [100], authors could utilize a time-filtered gating technique to separate the emission of single excitons and achieve pure single-photon signal. Continued advancements in experimental techniques and tailored system designs are crucial for achieving reproducible and efficient single-photon sources in plasmonic quantum systems.

### 2.3 Plasmon-exciton quantum emitters in strong coupling regime

In contrast to the weak coupling regime, both intermediate and strong light-matter coupling regimes bring about notable alterations not only in exciton relaxation rates and absorption efficiency but also in the quantum nature of transitions. In the strong coupling regime, describing light-matter coupling as a perturbation to Fermi's golden rule becomes inadequate, rendering the theoretical application of the Purcell effect impractical. This results in distinctive features specific to the strong coupling regime. Nonradiative processes may significantly decelerate in the strong coupling or even intermediate coupling regimes, thereby enhancing emitter excitation and decay via radiative channels [150,151]. On the other hand, the spontaneous emission rate no longer exhibits a linear dependence on the coupling strength, and achieving the ultrastrong coupling regime ($g > 0.1\omega_0$) can lead to emission rate saturation or reduction [152].

The most straightforward method to achieve strong coupling between semiconductor QEs and plasmonic structures involves using QE ensembles, allowing for the attainment of even the ultrastrong coupling regime [113,153–155]. Indeed, according to Eq. 4, an increase in the number of QEs coupled to the same cavity mode amplifies the collective dipole moment associated with the plasmon mode, resulting in a higher coupling strength $g$. However, the mere presence of more than one QE within the photon collection spot, constrained by the diffraction limit, alters the statistics of emitted photons, hindering the achievement of the desired generation mode for single photons or entangled pairs. Consequently, meeting the single-photon QE or entangled photon pair QE conditions necessitates the coupling of only one emitter with the plasmon mode. For the estimation of coupling strength for a single CQD interacting with a plasmonic cavity, it suggests that the transition dipole moment could be as high as 15 Debye [156]. Yet, achieving strong coupling for the case of a single QE is challenging, considering the typical measured damping rate of a single plasmon nanostructure is about 150-400 meV [104,157,158], making it difficult to reach $g$ values that meet the strict strong coupling criterion.

Santhosh et al. [110] achieved a coupling strength of 120 meV in the scattering spectrum by forming bowtie antennas with a single QD deposited inside the gap cavity. While this satisfies less strict criteria for strong light-matter coupling, the absence of emission measurements precludes the estimation of effects on QE properties. Subsequent work observed strong coupling between a single QD and a single gold nanorod via scattering spectrum measurements [157]. To maximize coupling strength, a core−shell QD was positioned below one end of the nanorod, forming a wedge nanogap cavity. This configuration achieved a substantial coupling strength of up to ∼114.34 meV (Rabi energy ∼234 meV), meeting the strict strong coupling criterion when considering the plasmon damping rate. Even without the nanogap, a Rabi splitting energy of ∼131 meV was observed, still meeting the less strict strong coupling criterion. However, no photon emission measurements were conducted in either study.

Compared to scattering, achieving strong coupling in the emission spectrum is more challenging due to the need to reduce the plasmonic mode volume, and thus, placing the QE in close proximity to the metallic surface. This proximity results in strong quenching of QE emission due to metal-induced energy and charge transfer. Despite these challenges, recent studies have demonstrated intermediate/strong coupling in the emission of single QDs using plasmonic nanocavities of complex geometries [39], such as plasmonic nanoslit (Figure 9a) [156], metal nanotips contacted with metal substrates (Figure 9b,c) [104,159], nanogap antenna (Figure 9d-f) [160], and bow-tie nanoantenna (Figure 9g-i) [116],. Leng et al. investigated a single CQDs in the gap between a gold nanoparticle and a silver film, achieving strong coupling in both scattering and emission spectra [160]. Based on the results of numerical simulations, the authors expected only one QD coupled to the nanogap mode. However, they also shown that more than one QD could be attached to the plasmon nanoparticle within employed synthesis procedure. For both scattering and emission spectrum the strong coupling regime was achieved with calculated coupling strength of 230 meV (115 meV in terms of this review) which is sufficient to meet the strong coupling criterion. In the article, the authors defined coupling strength as equal to the on-resonant vacuum Rabi energy, which is twice the definition of $g$ used in this review. By placing a single QD between the gold AFM tip and gold mirror, Park et al. observed a clear, high-contrast plexcitonic PL with a Rabi splitting of up to 163 meV in the emission spectrum [104]. This value meets the strong coupling criterion when compared with the plasmon damping mode of 126 meV and the exciton dumping of 97 meV. In further study by the same group, May et al. utilized the same experimental geometry [159]. Single-emitter coupling strength of 150 meV was achieved in this study by fitting the splitted PL spectrum. However, neither of these studies included photon correlation or time-resolved emission measurements, which are crucial for QE applications.

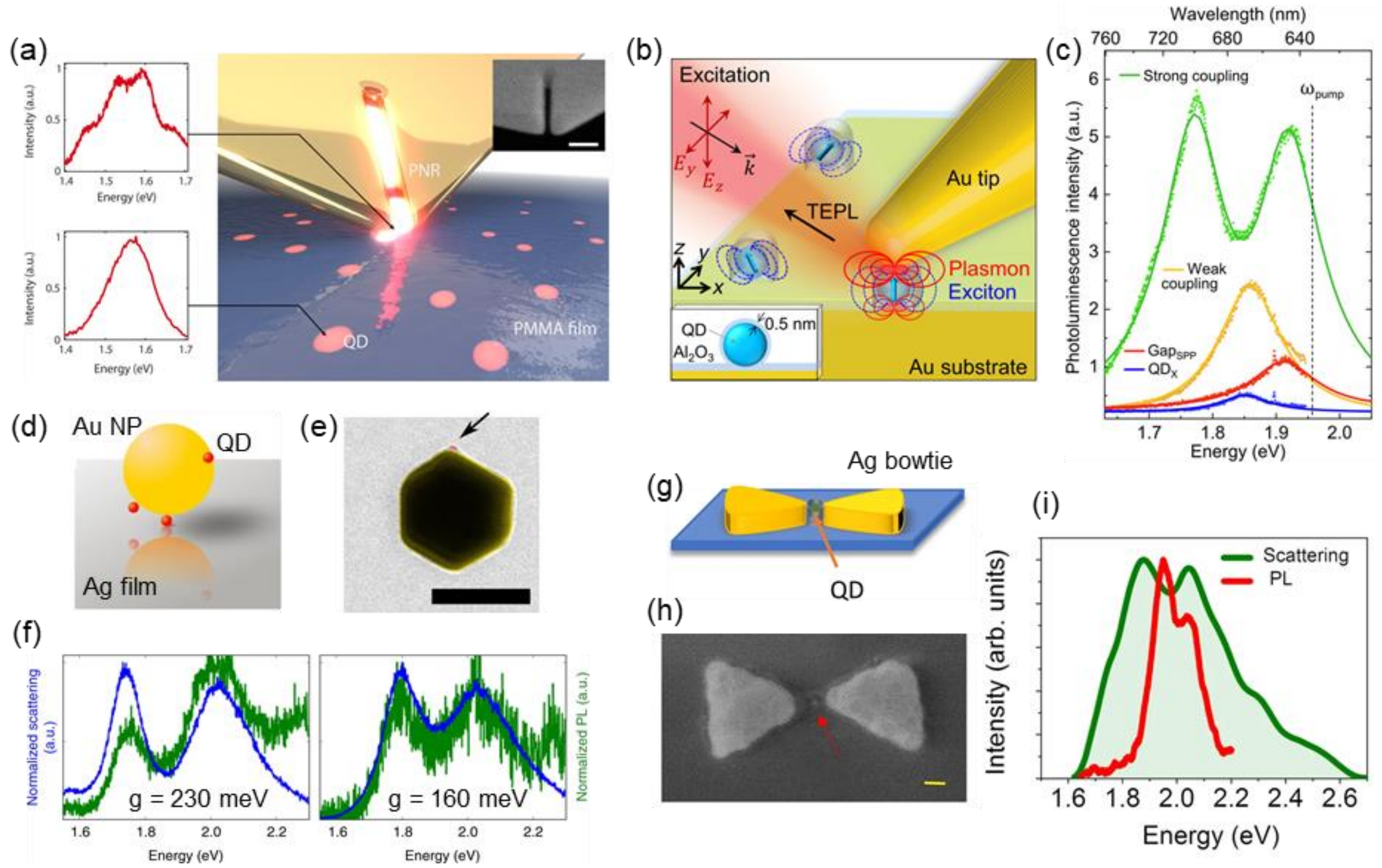

**Figure 9.** Examples of strong plasmon-exciton coupling obtained on a single-nanocrystal level. (a) Illustration of the plasmonic nanoresonator (PNR) probe interacting with single QDs embedded in a PMMA film. Left panel: The spectrum of a QD changes significantly when coupled to the slit-like PNR at the tip apex. Inset: SEM image of a nanoresonator at the apex of a probe tip. Scale bar is 100 nm. (b) Illustration of individual CdSe/ZnS QDs on an Au substrate and tilted Au tip inducing coupling between the plasmon and exciton. (c) PL spectra of the gap plasmon (red), QD exciton (blue), the weakly coupled plasmon-exciton mode (orange), and the strongly coupled plexciton mode (green) measured in an experimental system presented in (b). (d) Illustration of Au nanoparticle with QDs on its surface deposited on a silver film. (e) TEM image of single CdSe/CdS QD (marked by arrow) attached at the apex of a faceted Au nanoparticle. Scale bar is 100 nm. (f) Two examples of measured scattering spectra (blue) and PL spectra (green) for systems similar to presented on (d) and (e). Values of g parameters are calculated from fitting the measured spectra to coupled oscillator model. (g) Schematic of the plasmonic bowties with QD trapped in the nanogap. (h) STEM image of a device presented in (g) with one QD in the nanogap (red arrow). The scale bar is 20 nm. (i) Dark-field scattering spectrum (green) and PL spectrum (red) of the device presented in (h). (a) Reprinted from [156]. (b, c) Reprinted from [104]. (d, e, f) Reprinted from [160]. (g, h, i) Reprinted from [116].

Groß et al. achieved strong plasmon-exciton coupling between single CQDs and a slit nanocavity attached to an AFM cantilever. They confirmed the single-photon QE behavior of the CQD before coupling through $g^{(2)}(\tau)$ measurements [156]. The coupling strengths differed for the neutral exciton (110 meV) and charged exciton (44 meV), satisfying the strong coupling criterion considering the plasmon bandwidth of 78 meV.

Gupta et al. explored the strong plasmon-exciton coupling of single QDs to the plasmon gap mode of a gold bowtie antenna, obtaining coupling strengths between 52.5 meV and 110 meV in scattering spectra, meeting the less strict strong coupling criterion [116]. However, emission spectra indicated Rabi splitting values of only up to 110 meV, suggesting distinct microscopic mechanisms for scattering and emission spectra. Photon-correlation measurements demonstrated that some coupled systems met the single-photon QE conditions, while most did not, possibly due to the coupling of several QDs to the same cavity. Before coupling, the emission lifetime of single QDs averaged 24 ns, reducing by a factor of 5 after coupling to values ranging from 3 to 12 ns. This contrasts with the effects observed in the weak plasmon-exciton coupling regime, where plasmon-exciton emission lifetimes are orders of magnitude shorter than bare exciton lifetimes due to the strong Purcell effect. A similar effect was observed in transient absorption measurements with an ensemble of QDs strongly coupled to a plasmonic hole array, showing no substantial difference between the lifetime of the exciton state and the hybrid plasmon-exciton state [155].

## 3. Summary and outlook

The light-matter coupling between excitons of colloidal semiconductor QEs and plasmonic modes of metal nanostructures can significantly modify their emission properties. The strong improvement of the emission properties was demonstrated by many research, including emission enhancement with a factor up to 1900 and emission lifetime shortening down to 13 ps [118] with the possibility of total blinking suppression [44]. Two primary regimes have been utilized for this purpose: weak light-matter coupling, characterized by excitation enhancement and the Purcell effect, and strong light-matter coupling, leading to the alteration of quantum states within the system and the phenomenon of Rabi splitting. Various geometries of plasmon structures have been demonstrated for plasmon-exciton coupling (Figure 6 and Table 2).

Plexciton QEs based on luminescent colloidal nanocrystals have demonstrated the ability for both single-photon emission and multiphoton emission applications. Single-photon purity stands out as a crucial parameter for the design of single-photon QEs. In the absence of any filtering or signal processing, $g_0^{(2)}$ values as low as 0.2 were achieved (Table 2), and temporal filtering allowed for additional reduction down to 0.07 [123], approaching the average level achieved with uncoupled colloidal nanocrystals (Table 1). Reducing operating temperatures and applying spectral filtering are potential approaches to increase the single-photon purity of the plexciton emission. Studies have already demonstrated that different exciton lines can be observed and separated in the plasmon-exciton coupling mode at cryogenic temperatures [121]. However, almost only CQDs have been employed as individual QEs in these studies. In most recent studies PNC-based QEs were the sole means used for light-matter coupling with optical microcavities [161,162]. However in the recent study of Olejniczak et al. [76] authors directly demonstrated reversible plasmon-exciton coupling of single PNC-based QE with plasmonic metasurface and it's switching between single-photon and multiphoton emission modes. It should be noted that PNCs have shown strong potential as sources of single, indistinguishable photons [9], even more promising than many types of CQDs [48].

Another crucial aspect in the design of single-photon QEs pertains to the emission of indistinguishable photons. Indistinguishable photon emission occurs when the optical coherence time ($T_2$) of the QE approaches twice its radiative lifetime ($T_1$), meeting the transform limit given by $T_2 = 2T_1$. At cryogenic temperatures, the radiative lifetime of PNC emission can be as low as 180 ps [64], but this remains insufficient to meet the transform limit [9]. The Purcell effect has demonstrated a substantial reduction in the emission lifetime, with the maximum achieved value being approximately 540, by employing a nanogap nanoantenna with a silver nanocube on a gold mirror [118]. Extrapolating a similar order of Purcell factor to PNCs could potentially reduce the emission lifetime to a sub-picosecond scale. This reduction might facilitate the achievement of the transform limit even at temperatures higher than 6 K. For InP/ZnSe/ZnS CQDs, the lifetime must be reduced to the order of 125 ps to reach the transform limit. Given the initial lifetime of 15 ns for these CQDs, a Purcell factor of 120 is required to attain this reduced value. Two studies have achieved Purcell factors above 100 near the end of a gold nanocone [125] and gold nanocube [126] for the individual QE. However, it is noteworthy that single-photon purity was not conserved in these studies, necessitating the use of filtering techniques to achieve single-photon emission.

In turn, the achievement of multiphoton emission from colloidal semiconductor QEs has been widely asserted as a prerequisite for the generation of entangled photon pairs. However, only a high probability of photon pair emission from the same QE alone does not suffice for achieving quantum entanglement. The realization of quantum entanglement is a complex task that involves overcoming the FSS issue, which is inherent in all types of colloidal semiconductor QEs. Overcoming the FSS challenge is crucial for establishing quantum entanglement in these systems. Notably, the lowest values of FSS have been demonstrated for PNCs, with values reaching as low as ~200 μeV [80]. Utilizing Equation (3), we can estimate the exciton emission lifetime required to mitigate the effects of FSS. We assumed a spin dephasing time in QDs on the order of 10 ns [163,164], significantly longer than the lifetime of plasmon-accelerated exciton emission. However, the spin-flip scattering time ($\tau_{SS}$) for colloidal nanocrystals remains poorly investigated. We plotted diagrams illustrating the entanglement fidelity levels for various combinations of FSS and emission lifetimes, considering two significantly different values of $\tau_{SS}$: 10 ns (Figure 10a), characteristic of CdSe/ZnS epitaxial QDs [165], and 100 ps (Figure 10b), approximately the lower limit for total spin lifetime of colloidal PNCs at cryogenic temperatures [166]. It is evident that $\tau_{SS}$ predominantly influences the lowest fidelity levels. The most intriguing range of fidelity begins above 0.75 in both cases and corresponds to an emission lifetime of 3-4 ps for the FSS of ~200 μeV, which level indicated by the horizontal dashed lines in Figure 10. Notably, the lowest achieved value of plasmon-accelerated emission to date is 13 ps [118], as depicted by dashed vertical lines in Figure 10.

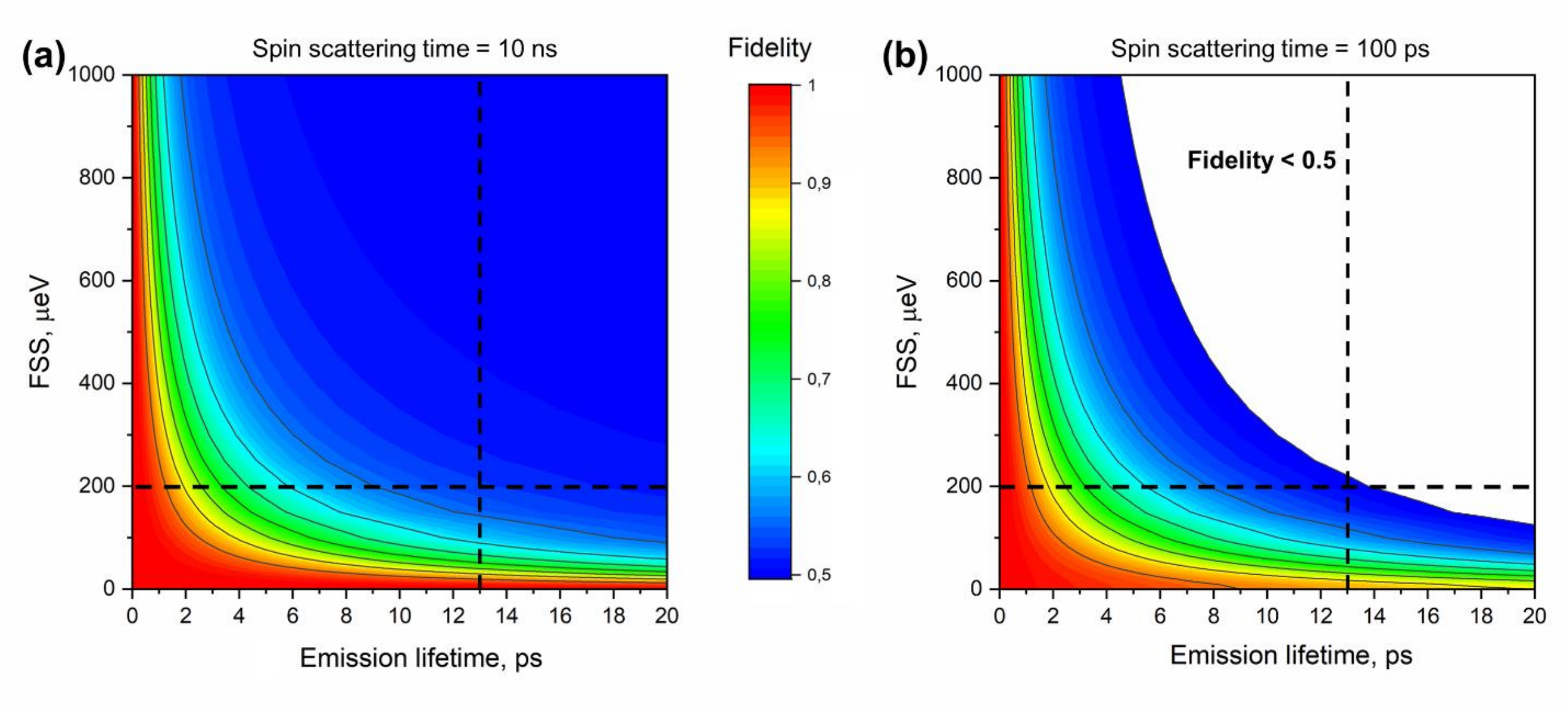

**Figure 10.** Diagrams showing the entanglement fidelity levels for different combinations of FSS and emission lifetime in QEs in the case of cross-dephasing time much longer than exciton emission lifetime ($\tau_{HV} \gg T_{1x}$) for two different values of spin-flip scattering time ($\tau_{SS}$) of (a) 10 ns, and (b) 100 ps. Horizontal dashed lines represent the minimal achieved value of FSS in colloidal nanocrystal QEs of ~200 μeV [80]. Vertical dashed lines represent the minimum achieved value of the emission lifetime of colloidal nanocrystal QEs of 13 ps [118]. White color on the panel (b) represents the values of entanglement fidelity below 0.5.

Thus, achieving the values of the exciton lifetime for the overcoming FSS issue necessitates further enhancement of the Purcell factor. The increase of the Purcell factor can be realized through an improvement of the quality factor and a reduction in the mode volume. Utilizing more sophisticated nanogap or dimer nanoantenna geometries can yield smaller mode volumes, thereby enabling the attainment of Purcell factors exceeding 5000 [41]. Even more promising technique involves coupling QEs with hybrid photonic-plasmonic cavities [167], which exhibit quality factors above 103 and normalized mode volumes down to $10^{-3}$, resulting in high Purcell factors on the order of $10^5$ [168]. The significant reduction of the mode volume with the increase in cavity quality factors also holds promise for achieving strong light-matter coupling with colloidal semiconductor QEs.

Recently the realization of strong plasmon-exciton coupling with a single QE has been clearly demonstrated in several studies. Rabi splitting in the scattering spectra of plexciton systems with values as high as 120 meV was observed in the geometry of a nanogap (bowtie dimer) nanoantenna [39]. Remarkably, also for the bowtie dimer nanoantenna, a highest Rabi splitting energy of up to 350 meV was achieved [116]. In the case of the emission spectrum, values of coupling strength up to 110 meV were attained using the nanoslit antenna [156]. However, despite these advancements, no significant improvement in the emission lifetime or intensity of QEs in this regime has been observed. It is noteworthy that only in the work of Gupta et al. [116] were time-resolved and photon correlation measurements performed, shedding light on the radiative dynamics of the system. Despite these contributions, the field remains relatively understudied, emphasizing the need for further investigation and exploration. Intriguing that an additional increase in the coupling strength beyond ~200 meV can lead to a situation where the description of the process by standard theoretical models, such as "Quantum Rabi" or "Jaynes–Cummings," becomes impossible. This regime is recognized as "ultrastrong coupling", characterized by a coupling strength to transition energy ratio exceeding 0.1. As the light–matter coupling strength enters the ultrastrong regime, it fundamentally alters the nature of the light and matter degrees of freedom. If the Rabi energy exceeds the exciton binding energy it leads to the formation of Wannier exciton-plasmon polariton, which may make possible a room-temperature Bose-Einstein condensation [169]. This transition opens new avenues for designing non-perturbatively coupled light–matter systems, likely leading to innovative applications [170]. To this moment, the values achieved for the coupling strength are close to the ultrastrong coupling threshold but still did not reach it, thus, the further improvement of the cavities is necessary to reach this intriguing regime.

In conclusion, the design of new plexciton QEs based on colloidal semiconductor nanocrystals represents an actively growing field that remains underexplored in several dimensions. The ongoing advancements in plasmonic cavities and light-matter coupling conditions hold the potential to unveil novel insights, particularly in the realm of ultrastrong coupling regimes. The quest for achieving ultrastrong coupling opens up opportunities for unprecedented discoveries, pushing the boundaries of our understanding in this relatively understudied domain. As parameters of plasmonic cavities continue to improve, new frontiers in ultrastrong coupling regimes await exploration, promising exciting findings that could reshape the approach in the design of QEs. Simultaneously, the pursuit of regimes enabling the emission of indistinguishable single photons and entangled photon pairs through the application of the Purcell effect marks a pivotal avenue. These achievements stand poised to revolutionize the applications of CQDs and PNCs in the realm of quantum technologies.

## Aknowledgements

The work was supported by MCIN and by the European Union NextGenerationEU/PRTR-C17.I1, as well as by IKUR Strategy under the collaboration agreement between Ikerbasque Foundation and Material Physics Center on behalf of the Department of Education of the Basque Government. YR acknowledges financial support from Grant TED2021-129457B-I00 funded by MCIN/AEI/10.13039/501100011033 and the European Union NextGenerationEU/PRTR and from the Department of Education of the Basque Government under project Ref. IT1526-22, and financial support from Gipuzkoa Quantum (Quan 22) funded by Gipuzkoa Provincial Council. YR and AO also acknowledge support from the Office of Naval Research Global (award no. N62909-22-1-2031). Authors express their gratitude to ChatGPT, an AI language model developed by OpenAI, for its valuable assistance in refining the language of this paper. Authors also thank Prof. Sergey I. Bozhevolnyi for sharing a figure.

## Supplementary material for

# Advancements and challenges in plasmon-exciton quantum emitters based on colloidal quantum dots

Adam Olejniczak,[1] Yury Rakovich,[1,2,3,4] Victor Krivenkov[1,4]*

[1]Centro de Física de Materiales (MPC, CSIC-UPV/EHU), Donostia - San Sebastián, 20018, Spain

[2]Donostia International Physics Center (DIPC), Donostia - San Sebastián, 20018, Spain

[3]Ikerbasque, Basque Foundation for Science, Bilbao, 48013, Spain

[4]Polymers and Materials: Physics, Chemistry and Technology, Chemistry Faculty, University of the Basque Country (UPV/EHU), Donostia - San Sebastián, 20018, Spain

*victor.krivenkov@ehu.eus

**List of contents:**

**Shortcuts:**

**FWHM** = full width at half maximum, **PL QY** = photoluminescence quantum yield, **X** = exciton, **XX** = biexciton, **RT** = room temperature, **C/S** = core/shell, **avg** = average, **PR** = plasmon resonance, **AWL** = amplitude weighted average lifetime, **R** = radiative decay, **NR** = nonradiative decay, **NP** = nanoparticle, **ML** = monolayer, **ZPL** = zero phonon line

Table S1A. Colloidal semiconductor quantum dots as quantum emitters.

| Material | Size, nm | Emission max | FWHM | $g^{(2)}(0)$ | PL QY, % | X lifetime, ns | XX lifetime, ns | Ref. |
|---|---|---|---|---|---|---|---|---|
| Core only | | | | | | | | |
| CdSe | - | 605 nm | - | 0.33 | 40-50 | 1.7 | - | [1] |
| Core-shell | | | | | | | | |
| CdSe/ZnS | 3.6 | 570 nm | - | 0.05 0.004 (background corrected) | 40 | 20 | - | [2] |
| | 3.6 | 575 nm | 35 nm | <0.03 | 40 | 20 | | [3] |
| CdSe/ZnS/CdS/ZnS | 4.5 | 560 nm | 39 nm | 0.12 | ~50 | 15-18 | 0.5 | [4,5] |
| CdSe/$Zn_{1-x}Cd_xS$ | - | ~ 635 nm | ~ 45 nm | <0.25 | - | 12 | | [6] |
| CdSe/CdS | 14 | - | - | - | - | 86 (bright state) | 0.7 | [7] |
| | 12.5 | 620 nm | 65.1 meV | 0.04 | - | 60 (exciton) | - | [8] |
| | - | - | - | - | 90 | 95.9 | 6.44 | [9] |
| | 10.6-13.1 | 655 nm | - | 0.11 (exciton) 0.2 (trion) | - | 125 (exciton) 42 (trion) | - | [10] |
| Wurtzite rod-in-sphere CdSe/CdS | 9.5-14.6 | 615 nm | 35 nm | 0.15 | 95 | 7-25 | - | [11] |
| Wurtzite CdSe/CdS | 17 | 650 nm | 40 nm | 0.07, 0.03 (after spectral filtering) | 50 | 638 | - | [12] |
| CdSe/CdS/SiO2 | 11 / 35 | 645 nm | 40 nm | 0.33 | 47 | 30 | - | [13] |
| CdSe/CdS/SiO2 | 11 / 90 | 635 nm | 40 nm | 0.06 | 34 | - | - | [14] |
| InP/ZnSe/ZnS | 10.5 | 625 nm | 5 µeV (4 K) | 0.13 (RT) 0.07 (4 K) | - | 3 (RT) 16.7 (4 K) | - | [15] |
| PbS/CdS | 6.5 (core) 9.3-17.2 (total) | 1280 nm and 1500 nm | 90 meV | 0.4 | 10-30 | 1000-3500 | - | [16] |
| InP/ZnSe | 10 | 629 nm | 40-80 meV | 0.03 | 70-100 | 21 | ~70 ps | [17] |
| CdSeTe/ZnS | 15 | 705 nm | ~100 meV (RT) | 0.06 0.01 (waveform engineering purification) | - | 138 | 2 | [18] |
| CdSeTe/ZnS | - | 705 nm | - | 0.32 0.04 (inside the waveguide) | - | - | - | [19] |
| CdSe/ $CdSe_xSe_{1-x}$/CdS quasi-type-II localization regime | 14 | - | - | - | - | 67 (bright state) | 1.5 | [7] |
| $CuInS_2$/ZnS | 12.5 | 1.93 eV | 128 meV (RT) | ~0.2 | 50 | 80-328 | - | [20] |
| Dot-in-rod | | | | | | | | |
| CdSe/CdS | 2.7 (CdSe core) | 579-605 nm | 100-140 meV (RT) | 0.075 | - | 46 | - | [21] |
| | 2.7 (CdSe core) | 600 nm | - | 0.11 (bright state), 0.47 (grey state) | 100 (bright state), 36 (grey state) | 65 (bright state), 11.6 (grey state) | - | [22] |
| | | 600 nm | | | | | | [23] |
| ZnSe/CdS | | 583 nm | | | | | | [23] |
| Dot-in-tetrapod | | | | | | | | |
| CdSe/CdS/PMMA | 4 (core) | 635 nm | ~110 meV (RT) | ~0.2 | 50 | 25 | - | [24] |

Table S1B. Colloidal perovskite nanocrystals as quantum emitters.

| Material | Size, nm | Emission max | FWHM | $g^{(2)}(0)$ | PL QY, % | X lifetime, ns | XX lifetime, ns | Ref. |
|---|---|---|---|---|---|---|---|---|
| All inorganic, core-only | | | | | | | | |
| $CsPb(Cl/Br)_3$ | 9.5 | - | 1 meV (6 K) | 0.3 | - | 0.18-0.25 | - | [25] |
| | 2-11 | 2.75 eV | 88 meV | 0.045 | 32 | 11 | 0.05 | [26] |
| $CsPbBr_3$ | 9.4 | 511 nm | 100 meV | 0.06 | 58 | 6.44 (RT) 0.355 (4 K) | - | [27] |
| | ~8.6 | 522 nm | - | - | 50-90 | 6.5 | 0.085 | [28] |
| | 2-11 | 2.52 eV | 110 meV | 0.02 | 84 | 5 | 0.022 | [26] |
| | 2-11 | 2.45 eV | 97 meV | 0.13 | 70 | 9 | 0.22 | [26] |
| | | 2.4 eV | 110 meV | - | - | 4.8 | - | [29] |
| | 7.7 | 512 nm | | 0.1 | | 6 | 0.04 | [30] |
| | 10 | 515 nm | - | ~0.15 | 50 | 8.3 | 3.3 (rad) 0.25 – 1 (Auger) | [31] |
| | 11-15 | - | - | - | 51.6 | - | 0.074-0.15 | [32] |
| | 3.5-8.5 | 473-506 nm | - | - | 34.7-78 | - | 3.1 - 51.5 ps | [33] |
| | 13.5 | ~520 nm | 17-27 μeV (3.6 K) | < 0.04 (3.6 K) | 70.8 (RT) 95.6 (5 K) | 0.21-0.27 | - | [34] |
| | 4.5 (ultrasmall) 7 (small) 14 (large) | 2.5-2.59 eV (ultrasmall) 2.46 eV (small) 2.42 eV (large) | 85-127 110 meV (small) 95 meV (large) | 0.18 (small) 0.32 (large) | - | - | - | [35] |
| | 10 | 2.419 eV | - | 0.34 (avg) | - | - | - | [36] |
| | 10 | 505.2 nm | 72.19 | 0.05 | 63-73 | 5 | | [37] |
| | 5-25 | 490-515 nm | - | 0.07 (7.1 nm) 0.83 (21.1 nm) 0.05 (5 nm) | - | < 4 | - | [38] |
| | 19.8 | 2.39 eV (RT) | 74.5 meV (RT) | 0.019-0.052 (4 K) | - | 0.186-0.225 (4 K) | - | [39] |
| $CsPb(Br/I)_3$ | 15 | 520 nm | ~15 nm | 0.2 | <25% | 6.3 | - | [40] |
| $CsPbI_3$ | 9.3-11.2 | ~ 680 nm | 25.1 nm | 0.05 | 40 - 50 | 13.2-15.1 | 0.093 | [41] |
| | 11.2 | ~ 680 nm | 90 meV | - | 41 | 43.4 | - | [42] |
| | - | 685 nm | - | - | 100 | 36 | - | [43] |
| | 2-11 | 2.02 eV | 160 | 0.08 | 80 | 22 | 0.27 | [26] |
| | 2-11 | 1.92 eV | 120 | 0.13 | 70 | 17 | 0.44 | [26] |
| | 2-11 | 1.84 eV | 105 | 0.54 | 40 | 55 | 4.2 | [26] |
| | 9.3 | 700 nm | - | ~0.05 (4 K) | - | 0.93 / 1.02 (4 K) | ~0.41 (4 K) | [44] |
| | 9.3 | 690 nm | ~200 μeV (4 K) | ~0.04 (RT) | - | 46.9 (RT) 1.04 (4 K) | 2.9 (RT) | [45] |
| | 9.3 | 701 nm | 137 μeV (4 K) | - | - | 1.5 (4 K) | - | [46] |
| | 6.6 | 1.895 eV | 70 to 140 meV | >0.02 (avg 0.096) | - | - | - | [36] |
| | 10 | 1.816 eV | <60 meV | >0.085 (avg 0.18) | - | - | - | [36] |
| | 5 | 495 nm | 85 meV (RT), 2 meV (4 K) | 0.27 (RT) | - | 2.6 (RT) 0.54 (4 K) | - | [47] |
| | 11.2 | - | - | - | 65 (RT) | 1 (4 K) | - | [48] |
| | 9.8 | - | - | 0.021 (0.061 avg) | 95 | 9.5 | - | [49] |
| All inorganic, core-shell | | | | | | | | |
| $CsPbBr_3$/ $CsCaBr_3$ | 7 (small) 14 (large) | 2.52 eV | Down to 35 meV (both) | 0.1 (small) 0.16 (large) | - | ~5 | - | [35] |
| $CsPbBr_3$/CdS | 14.6 (core) 21.3 (C/S) | 531 nm | 20 nm | 0.43 (C/S) | 88 (C/S) | 22.8 (C/S) | - | [50] |
| Hybrid organic-inorganic, core-only | | | | | | | | |
| $FAPbBr_3$ | 6-14 | - | - | - | 91.2 | - | ~0.15-0.5 | [32] |
| | 11 | 524 nm | 21.3 nm | 0.063 | 75 | 13.8 | 0.15 | [51] |
| | 11 | - | - | <0.1 | - | 19.5-23.9 | - | [52] |
| | 10 | 523.5 nm | 87.56 meV | 0.013 | 63-73 | 36 | - | [37] |
| | 10.5 | 530 nm | - | ~0.12 | - | 20.6 | 0.085 | [53] |
| $Cs_{0.2}FA_{0.8}PbBr_3$ | 11 | 517.1 nm | 79.67 meV | 0.04 | 63-73 | ~15.5 (avg) | - | [37] |
| $MAPbI_3$ | 11 | - | - | - | 47.6 | - | ~0.2-0.4 | [32] |

Table S1C. Semiconductor 2D materials as quantum emitters.

| Material | Temperature, K | Emission max | FWHM | $g^{(2)}(0)$ | PL QY | X lifetime, ns | XX lifetime | Ref. |
|---|---|---|---|---|---|---|---|---|
| $WSe_2$ | 4 | 782.72 nm | 60 µeV | 0.17 | - | 4.14 | - | [54] |
| | 4.2 | 1.68 eV | | 0.36 | - | | - | [55] |
| | 4.2 | 1.71-1.73 eV | Down to 112 µeV<br>130 µeV (avg) | 0.14 | - | 1.79 | - | [56] |
| | 15-20 | 1.63-1.75 eV | >100 µeV | <0.35 | - | 0.6 | - | [57] |
| | 10 | 1.707 eV | ~120 µeV | 0.28 (0.03 background corrected) | - | 1.8 | - | [58] |
| | 4.2 | 1.68 eV | 110 µeV | 0.18 | - | 1.51 | - | [59] |
| | 4.2 | 1.7167 / 1.7206 eV | | 0.397 / 0.286 | - | 0.793 / 1.504 | - | [60] |
| | 10 | 750-780 nm | 180 µeV | 0.087-0.18 | - | 3.08-8.81 | - | [61] |
| | 3.5 | 775-835 nm<br>(1 ML) | <200 µeV | 0.07 | - | 2.8 | - | [62] |
| | 3.5 | 801.08 nm<br>(2 ML) | 131 µeV | 0.03 | - | 4.8 | - | [62] |
| | 5 | ~1.67 eV | 230 µeV | 0.13 | - | 1 | - | [63] |
| | 4 | | | 0.3 | - | | - | [64] |
| | | 807 nm | | 0.133 (0.034 with temporal filtering) | - | | - | [65] |
| | 3.8 | 1.6 eV | down to 170 µeV | 0.2 | - | 3.8 | - | [66] |
| | 10 | 750-770 nm | ~1.7 meV | 0.29 | - | 9.4 | - | [67] |
| $WS_2$ | 10 | 640 nm | 12.1 meV | 0.31 | - | 1.4 | - | [67] |
| $MoS_2$ | 5 | 1.75 eV | 28 meV (10 K) | 0.23 | - | 1.73 µs | - | [68] |
| $MoSe_2$ | 3.8 | 1.584 eV | 160 µeV | 0.28 | - | 0.19 | - | [69] |
| | 4 | 1.57-1.67 | 150-400 µeV | | - | | - | [70] |
| | | 1.58-1.64 eV | 200-500 µeV | | - | | - | [71] |
| $MoTe_2$ | 11 | 1.124 eV | 980 µeV | 0.058 | - | 22.2 | - | [72] |
| $WO_3$ | RT | 730 nm | <50 nm | <0.4 | - | 3.5 | - | [73] |
| GaSe | 10 | 1.96 eV | 3.7 meV | 0.33 | - | 7 | - | [74] |

Table S2A. Plasmon cavity – quantum emitter coupled systems with colloidal semiconductor quantum dots and perovskite nanocrystals.

| Cavity geometry | PR, nm | Quantum emitter | PL max, nm | Uncoupled / Coupled | | PL increase | Purcell factor | Ref. |
|---|---|---|---|---|---|---|---|---|
| | | | | Lifetime, ns | $g^{(2)}(0)$ | | | |
| **Nanogap nanoantennas** | | | | | | | | |
| Deformed Au bowtie (gap: 22 nm) | 800 | CdSe/ZnS | 655 | 35 / 19 | 0.38 / 0.41 | 7 | - | [75] |
| Parallel Au nanobars (gap: 55 nm) | 660 | thick shell CdSe/CdS | 640 | 40-80 / 5-45 | ~0.15 / 0.3-0.7 | - | 1.4-9.6 | [76] |
| Au nanorods linear dimer (gap: ~35 nm) | ~620 | CdSe/ZnS | 620 | 14.2 / 5 | 0.25 / 0.5-0.65 | 1.6-2.2 | - | [77] |
| Ag bowtie (gap: 20 nm) | 636 | CdSe/ZnS | 633 | 22 / ~6-7 | 0.25-0.4 / 0.5-0.8 | - | - | [78] |
| Au NP dimer (gap: 14 nm) | 660 | CdSe/CdS/ZnS | 660 | 2.0 / 0.18 | - | 1.3 | 11.1 | [79] |
| Au nanorods assembled into U-shape, gap 1: 29 nm | 715 | $CdSeTe/ZnS@SiO_2$ (10 nm silica shell) | 808 | 291 / 6.47 | - | - | 45 | [80] |
| Au nanorods assembled into U-shape, gap 2: 26 nm | | | | 248 / 1.88 | | | 132 | |
| Parallel Au nanorods (gap: 36 nm) | 645 | $CdSeTe/ZnS@SiO_2$ (10 nm silica shell) | 808 | 246 / 35 | - | - | 7 | [81] |
| Au NP dimer in solution (gap: 5 nm) | 590 | CdSeS/ZnS | 630 | - / 0.065 | - | 30 | - | [82] |
| **Nanopatch antennas** | | | | | | | | |
| Ag nanocube / Au film (gap: ~10 nm) | 630 | CdSe/ZnS | 630 | 6.8 / 0.013 | 0.17 / 0.32 | 1900 | 540 | [83] |
| Ag nanocube / Au film (gap: 15nm, PMMA layer) | 560 | CdZe/ZnS | 655 | 25 / 1 | 0.1 (avg.) / 0.7-1.0 | ~4 | ~70 | [84] |
| Au nanodisk / Au film (gap: 45 nm, PMMA layer) | - | thick shell CdSe/CdS | 635 | 36 / < 1.5 | 0.2-0.3 / 1 | 72 | 72 | [85] |
| **Single nanoantenna** | | | | | | | | |
| Ag covered AFM tip (coupled @ 6 nm, approach up to 2 nm) *(reversible coupling)* | 450 | CdSe/ZnS | 610 | 26.7 / 0.4 / 17.06 (AWL) | 0.09 / 0.85- 0.88 / 0.08 | <2.5 | - | [86] |
| Au nanocone (QD on an AFM tip approached to the antenna) | 625 | thick shell CdSe/CdS | 650 | X: 62 / 1.6 XX: 4 / 0.5 | 0.30 / 1.18 | ≤45 | X: ~109 XX: ~100 | [87] |
| Au nanocube (placed close to QD by AFM manipulation) *(reversible coupling)* | 600 | CdSe/ZnS | 610 | 29.6 / 0.4 / 17.4 (AWL) | 0.14 / 0.97 / 0.13 | 1.3 | R: 103 NR: 39 | [88] |
| Au NP (placed close to QD by AFM manipulation) *(reversible coupling)* | 530 | CdSe/ZnS | 635 | 34 / 1.7 / 27 | - | ~1.2 | R: 7.2 NR: 291 | [89] |
| Au nanocube@resin (~20 nm photo-polymerized resin layer with QDs) | 680 | CdSe/CdS/ZnS | 620 | 17.5 / 0.63 | 0.2-0.35 / 0.35 | - | 24 | [90] |
| Au $NPs@SiO_2$ (7 nm silica shell) | 540 | core/multishell CdSe/CdS/ZnS | 580 | 21 / 5.4 | - | 2.4 | ~3.9 | [91] |
| Ag $NPs@SiO_2$ (15 nm silica shell) | 439-446 | CdSe/ZnS | 605 | 7.72 / 0.56 (AWL) | 0.09 / 0.81 | ~2 | - | [92] |
| Ag nanoprisms (10 nm PMMA layer) | 700 | CdSe/ZnS | 600 | 25 / 5 | - / 0.2 | R: 12.5 NR: 2.5 | - | [93] |
| Ag nanowire (50 nm PMMA+QDs layer) | - | CdSe/ZnS | 650 | 25.4 / 3.2 | - / 0.3 | - | 17.3 | [94] |
| Ag nanowire (10 nm $Al_2O_3$ layer) | - | CdSe/ZnS | 650 | 24.2 / 5.6 | - / 0.29 | - | - | [95] |
| Ag nanowire (30 nm PMMA layer) | - | CdSe/ZnS | 655 | 22 / 13 | - / <0.3 | - | ~1.6-2.5 | [96] |
| Ag nanowire (distance 30-200 nm, QD placement by microfluidic system) | - | CdSe/ZnS | 620 | 17 / 8 | - | - | ~2 | [97] |
| **QE-in-shell** | | | | | | | | |
| Au shell (QD inside spherical Au nanoshell resonator) | 750 | thick shell $CdSe/CdS@SiO_2$ (35 nm silica shell) | 670 | 123 / 20 | 0.2-0.28 / 0.74-0.90 | 1.12 | ~6 | [98] |
| **Bullseye antennas** | | | | | | | | |
| Au bullseye | 617 | CdSe/CdS@PMMA (25 nm PMMA shell) | 2.005 eV | 1.63 / <0.23 | - / <0.5 (~0.4) | 5.5 | 7 | [99] |
| | | | 2.004 eV | 60.5 / 2.2 | | | 28 | |
| Ag bullseye (250 nm $SiO_2$ layer, 100 nm PMMA layer) | - | CdTe/ZnS | 780 | 170 / - | 0.12 / 0.37 | - | - | [100] |
| Au bullseye (174 nm PMMA layer, 5 nm $Al_2O_3$ layer) | 660-680 | thick shell CdSe/CdS | ~660 | X: 67 / - BX: 11/ - | - / 0.68 (0.07 filtered) | - | - | [101] |
| **Ordered nanoantenna arrays / metasurfaces** | | | | | | | | |
| Ag NP square array | 582 | CdSe/ZnS | 570 | 29 / 6.8 (AWL) | 0.02 / 0.34 | 1.5 | 6 | [102] |
| Ag NP square array (covered with PDMS) | 510 | | | 24 / 1.67 (AWL) | 0.018 / 0.27 | 4.2 | 22 | |
| Ag 1D grating (20 nm $SiO_2$ layer, < 5 nm PMMA layer) | - | CdSe/ZnS | 600 | 17.1 / 0.7-7.3 | 0.06 / 0.21-0.25 | 3-5.3 | R: 3.6 NR: ~30 | [103] |
| Ag NP square array, period: 400 nm | 620 | CdSe/ZnS | 620 | 22-25 / 0.39-0.95 | - | - | 17-60 | [104] |
| Ag NP square array, period: 450 nm | 690 | | | 22-25 / 0.38-1.34 | | | 25-80 | |
| **Disordered nanoantenna arrays / metasurfaces** | | | | | | | | |
| Ag nanocube metasurface (on top of ~15 nm PMMA+QDs layer) *(reversible coupling)* | 515 | $CsPbBr_{2.5}I_{0.5}$ | 520 | 0.8-1.1 / 0.4 | <0.5 / >0.5 | 2.2-7.5 | 5-12 | [40] |
| | | thick shell CdSe/CdS | 620 | 5.8 / 0.6 | <0.5 / <0.5 | ~0.24 | - | |
| | | core/multishell CdSe/ZnS/CdS/ZnS | 560 | 3.2 / 0.5 | <0.5 / >0.5 | ~1.3 | - | |
| Ag nanoprisms (0-135 nm PMMA layer) | 570 | core/multishell CdSe/ZnS/CdS/ZnS | 560 | 3.7 / 0.3 – 0.5 | 0.8-1.3 | 0.1-1.9 | 23-37 | [5] |
| Ag NPs (0-135 nm PMMA layer) | 430 | | | 3.7 / 0.4 | 0.9 – 1.3 | ~1.5 (avg.) | - | |
| Au nanorods (10-170 nm PMMA layer) | 610 | core/multishell CdSe/ZnS/CdS/ZnS | 560 | 18 / 1.5 | 0.1-0.3 / 0.712-0.968 | ~0.33 | ~3 | [4] |
| Au NPs@$SiO_2$ (10 nm silica shell) | 612 | CdSe/CdS | 629 | 37.93 / 4.03-22.67 (AWL) | 0.1-0.2 / 0.6-0.8 | - | X: 1.4 XX: 5.6 | [105] |
| Ag NPs (on top of 25 nm PMMA+QDs layer) | 430 | CdSe/ZnS | 610 | 20-30 / 0-10 | 0.05-0.2 / 0.05-0.5 | 2.5-4 | R: 2-10 NR: <10 | [106] |
| Ag nanoflakes film | - | thick shell CdSe/CdS | 620 | 39 / 0.9 | 0.17 / 0.9 | 0.5 | R: 4 NR: 60-280 | [107] |
| Au NPs (20 nm $Al_2O_3$ layer) | 590 | CdSe/CdS | 638 | - | 0.021-0.045 / 1.16 | 2.5-11.9 | - | [108] |
| Au NPs (on top of 23 nm PMMA+QDs layer) | 530 | CdZnSe/ZnS | 532 | 15 / 2 | 0.13 / 0.69 | 2-6 | 2-5 | [109] |
| Au NPs@$SiO_2$ (5 nm silica shell) (isolated NPs on glass) | 526 | core/multishell CdSe/CdS/ ZnCdS/ZnS | 600 | 21.7 / 14 | - | - | - | [110] |
| Au NPs@$SiO_2$ (5 nm silica shell) (aggregates of NPs on glass) | 590 | | | 21.7 / 1.1 | - | 3 | ~10 | |
| **Au films** | | | | | | | | |
| Au film (on top of 150 nm PMMA+QDs layer) | - | CdSe/ZnS | 617 | ~18 / 6-8 | 0 / ≤0.45 | - | X: 1.5 XX: 10 | [111] |
| Au film | ~650 | thick shell CdSe/CdS | 660 | 75 / 0.8-2.9 | ~0 / ~1 | 2-7 | 14-60 | [112] |
| Au film (30 nm $SiO_2$ layer) | ~650 | CdSe/CdS | 620 | 60 / 6-40 | - / <0.2 | 1.5-10 | 10 | [113] |

Table S2B. Plasmon cavity – quantum emitter coupled systems with 2D materials.

| Cavity geometry | | PR, nm | Quantum emitter | PL max, nm | Uncoupled / Coupled | | PL increase | Purcell factor | Ref. |
|---|---|---|---|---|---|---|---|---|---|
| | | | | | Lifetime, ns | $g^{(2)}(0)$ | | | |
| **Nanogap nanoantennas** | | | | | | | | | |
| Two parallel Au bars (gap: 96 nm) | | - | $WSe_2$ ML (strained over the gap between metal bars) | 742-826 | 14 / 1.07 | - / 0.42 | - | 1.9-15 | [114] |
| Linear Au bars (gap: 100 nm) | | - | $WSe_2$ ML (strained over the gap between metal bars) | 720-775 | - / 1.9-2.8 | - / 0.19-0.23 | - | - | [115] |
| **Single nanoantenna** | | | | | | | | | |
| Au nanostars on top of a ML (tip size: 2-3 nm, 2 nm bilayer of buffer molecules on NP surface) | | 795 | $WSe_2$ ML | 750 | 15.1 / 5.5 | 0.23 / 0.23 | 1.4 | 2.7 | [116] |
| Ag nanowire | | - | WSe2 ML (strained on top of a nanowire) | 720-770 | - / 5.0 | - / 0.1 | - | - | [117] |
| Au NPs (attached to hBN flakes using AFM) | 1 Au NP | 535 | point defects in hBN | 578 (ZPL) | 4.92 / 2.68 | 0.24 / 0.26 | 1.55 | 1.89 | [118] |
| | 2 Au NPs | | | | 4.92 / 1.54 | 0.24 / 0.31 | 3.10 | 3.19 | |
| Square Al nanoantenna (scanning 10-20 nm above the surface) | | 620 | point defects in hBN | 620 | 3.5 / 1.6 | < 0.5 / - | - | 2.2 | [119] |
| **Metal film** | | | | | | | | | |
| Rough Ag film (3 nm $Al_2O_3$ layer) | | 720-800 | $WSe_2$ ML (strained on surface roughness) | 780-840 | 4-6 / 0.1-0.5 | - / 0.25 | - | - | [120] |
| Rough metal layer | Ag | - | $MoS_2$ ML (strained on surface roughness) | 700 | - | - / 0.88 | - | - | [121] |
| | Au | | | | - | - / 0.42 | | | |
| **Nanopatch antenna** | | | | | | | | | |
| Au cube / Au film (gap: 5 nm) (2 x 2 nm $Al_2O_3$ layer) | | 750-800 | $WSe_2$ ML | 750-800 | 4.0 / 0.266 | 0.22 / 0.21 | 13 | 181 | [122] |
| **Ordered nanoparticle array** | | | | | | | | | |
| Square array of metal NPs | Ag | 640 | point defects in hBN (flakes on top of NP array) | 600 | $t_1$: 0.63 / 0.42 $t_2$: 3.4 / 2.6 | 0.02 / 0.04 | ~2.5 | - | [123] |
| | Au | 650 | | 652 | 4 / 0.6 | 0.23 / 0.47 | - | - | |
| | $Ag/Al_2O_3$ (5 nm $Al_2O_3$) | 638 | | 680 | 8.5 / 0.27 | 0.06 / 0.29 | - | ~30 | |
| Square array of $Si/Au/Al_2O_3$ nanopillars (6 nm $Al_2O_3$ layer) | | 740-880 | $WSe_2$ ML (strained on individual nanopillars) | 730-770 | 5.2 / 2.2 | - / 0.3 | 2 | 2.4 | [124] |